\documentclass[prb,twocolumn,showpacs,amsmath,amssymb,floatfix,superscriptaddress]{revtex4-2}

\usepackage{graphicx}
\usepackage[usenames,dvipsnames]{color} 
\usepackage{bm} 
\usepackage[update,prepend]{epstopdf}
\usepackage{soul} 
\usepackage{dcolumn} 
\usepackage{multirow}
\usepackage{tabularx}
\usepackage{amssymb}
\usepackage{tikz}

\newcommand{\rfive}{$\sqrt{5}\times\sqrt{5}$}
\newcommand{\dt}{DyTe$_2$}

\begin{document}

\title{Unveiling the Electronic, Transport and Migration properties \\ of the Te-defect lattice in DyTe$_{1.8}$}
	
\author{Jinwoong Kim}
\email{jinwoong.kim@csun.edu}
\author{Nicholas Kioussis}
\email{nick.kioussis@csun.edu}
	
\affiliation{
	Department of Physics and Astronomy,
	California State University,
	Northridge, California 91330, USA
}

\date{\today}
\begin{abstract}
The rare-earth ditellurides are known to form two-dimensional
square lattice where the strong Fermi surface nesting leads to 
structural modulation. In contrast to charge density waves,
the supercell modulation is accompanied 
by the formation of the periodic Te vacancy network, 
where the Te deficiency affects the nesting vector (i.e. the supercell size) via tuning the chemical potential.
In this work, first principles electronic structure calculations for the {\rfive} supercell,
that commonly appears in this family of tellurides, unveil interesting electronic, transport, and migration properties of the Te defect lattice in DyTe$_{1.8}$.
The reconstruction of the Te-deficient square lattice, consisting of a single Te-dimer and a pair Te-trimers per unit cell, gives rise to an out-of-plane polarization, whose direction depends 
on the position of the dimer.
This results in various close-in-energy
parallel and antiparallel polarization configurations
of successive Te layers depending on the dimer positions.
We predict 
that the orientation of the Te dimers,
and hence the corresponding structural motifs, can be reversibly switched between two in-plane perpendicular directions under tensile epitaxial strain via a piezoelectric substrate, resulting in a colossal conductivity switching.
Furthermore, the Te-dimer orientations result in asymmetric Fermi surface which can be confirmed by quantum oscillations measurements. 
Finally, we present numerical results for the migration paths and energy landscape through various divacancy configurations in the presence or absence of epitaxial strain.
\end{abstract}
	
\maketitle

\section{Introduction}

The rare-earth ditellurides $R$Te$_2$ ($R$ represents a lanthanide element) are a family of layered materials renowned for a wide range of intriguing properties at low-temperature.
These include charge density wave (CDW)\cite{Fisher2005,lanzara2007} magnetism,\cite{Kwon2000} superconductivity under high pressure,\cite{Takabatake2003} as well as the simultaneous presence of CDW, superconductivity, and antiferromagnetism.\cite{Fisher2005}
The unit cell of the crystal structure comprises of double-corrugated $R$-Te quasi-ionic layers separated by single planar square Te sheets stacked along the $[001]$ direction, as illustrated in Fig.~\ref{fig:nesting}(a). 
From the electronic structure point of view, each ($R$-Te)$^{1+}$ pair 
on the corrugated layer donates one electron  to the Te atoms (Te$_s$) in the square plane.\cite{Hoffmann1987,Kikuchi1998}

In contrast to the rare earth tritellurides, $R$Te$_3$, which form stoichiometric compounds, the $R$Te$_{2-\delta}$ ($0.0 \lesssim \delta \lesssim 0.2$) compounds 
host Te$_s$ vacancies on the Te$_s$ square layer, giving rise to a wide range of structural motifs which depends sensitively  on the Te deficiency $\delta$ and the rare earth cation.\cite{Falson2024,Poddig2021} 
The Te$_s$ vacancies effectively contribute extra charge to the layer which along with the isolated Te$^{2-}$ anions form differently ordered patterns within the Te sheets. 
This results in superstructure modulations which can be commensurate with the parent structure, such as in CeSe$_{1.9}$\cite{Plambeck-Fischer1989} and Gd$_8$Se$_{15}$\cite{Dashav2000} structures, and 
incommensurate modulated structures such as DySe$_{1.84}$\cite{DerLee1997} and $R$Se$_{1.84}$ ($R$=La-Nd, Sm)\cite{Graf2009}.

Recently, we have reported\cite{Falson2024} experimental 
and first principles electronic structure studies of the supercell formation in epitaxial DyTe$_{2-\delta}$ thin films, where for $\delta$=0.2 the modulation is 
a ($\sqrt{5}\times\sqrt{5}$) R26.6$^\circ$~$\times$~2 superlattice. 
The superlattice emerges due to a periodic Te-defect lattice, which also acts to open a gap in the electronic spectrum and induces semiconducting transport behavior. First principles calculations point towards nesting conditions of the Fermi surface at a $\bm{q}$-vector that corresponds to the \rfive~condition, suggesting that the formation of the defect lattice results from a similar driving force to the conventional picture of CDW formation, where sections of the Fermi surface are gapped out by the formation of supercells with periodicity corresponding to the nesting condition\cite{Lee1996}.
Our total energy calculations\cite{Falson2024} of the \rfive ~supercell with Te$_s$ mono- and divacancies in different configurations for $\delta$ = 0.1 and 0.2, showed that the second nearest-neighbor (SNN) vacancy configuration (denoted as the A-C configuration in Fig. 5c of Ref. \cite{Falson2024}) 
has the lowest formation energy among the various considered structures, in agreement with previous 
X-ray diffraction measurements in SmTe$_{1.8}$.\cite{Ijjaali2006} 
The SNN di-vacancy configuration shown in Fig.~\ref{fig:structure}(c), where the di-vacancies nucleate on sites 0 and 4 $\{0,4\}$ in 
Fig.~\ref{fig:structure}(a), consists of an ordered lattice of Te$_s$ dimers and trimers. 

Furthermore, first principles calculations of the ($\sqrt{5}\times\sqrt{5}$)$\times$~2 modulation along the $c$ axis corresponding to the nucleation
of  a pair of A-C (0,4) divacancies on two Te square layers, reveals that the SNN A-C divacancy appears at a laterally displaced position $(0.5, 0.5)$, relative to the divacancy on the adjacent Te square net layer. 
\cite{Falson2024}

The objective of this work is to unveil the emergence of novel interesting properties of the periodic Te-defect lattice associated with the onset of ferroelectric polarization which depends sensitively on the position of the Te-dimer.
In Sec. II we outline the methodology employed. In Sec. III we present results of (a) the position and orientation of the various Te-dimer configurations, (b) their polarization properties,  (c) the electronic structure, (d) the Fermi surface anisotropy, (e) and (f) the  transport and migration properties in the absence or presence of epitaxial strain. Conclusions are summarized in Sev. IV.


\section{Methodology}
The density functional theory calculations are performed using the Vienna {\it ab intio} simulation package ({\small VASP})\cite{Kresse1996,Kresse1996b} 
with the projector augmented wave method.\cite{Blochl1994,Kresse1999} The Dy$–4f$ states are treated as core with valence configurations $5p^65d^16s^2$ and $5s^25p^4$ for Te.
All structures are optimized using the PBEsol exchange correlation functional,\cite{Perdew2008} and the Bloch states are calculated and Wannierized\cite{Mostofi2014,Marzari2012} with the modified Becke-Johnson potential\cite{Becke2006,Tran2009} which provides accurate band gaps, effective masses, and frontier-band ordering.
The momentum space is sampled at a $80 \times 80 \times 80$
k-point mesh with the Wannier interpolation scheme.
The conductivity is calculated using the semi-classical Boltzmann 
transport theory assuming a constant relaxation time.
The tetrahedron method is employed to obtain the Fermi surface 
and its cross section area.
In the polarization calculation, the topmost 51 valence bands 
are Wannierized for each spin channel where three and two 
$p$ orbitals are initially projected on the Te$_c$ and Te$_s$ atoms, respectively, and five $s$ orbitals are projected at the 
center of the Te-Te bonds
in the square lattice. The integer ionic charges $Z_n$ 
in Eq.~\ref{eq:pol} are $+3$ for Dy and $+4$ for Te atoms, respectively.

\section{RESULTS AND DISCUSSION}

\subsection{Crystal structure} \label{sec:str}
\begin{figure} [t]
	\centering
	\includegraphics[width=8.6cm]{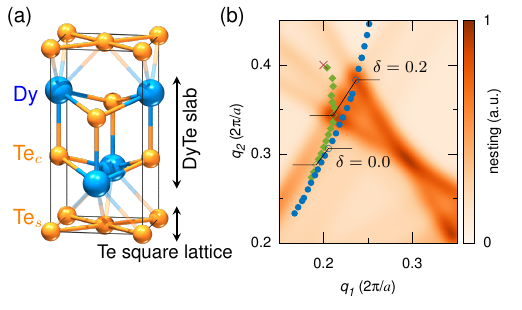}
	\caption{(a) Unit cell of pristine DyTe$_2$ with two Te sites (Te$_s$) on the square net layer. (b) Calculated nesting
		function of DyTe$_2$. The back ground color map illustrates Gaussian-smeared nesting function for an
		electron doping level corresponding to Te deficiency,
		$\delta=0.2$. The cross point indicates {\rfive} wave
		vector. The two local nesting function peaks are marked
		with dots upon the doping level, that moves away from the
		origin with increasing electron doping.
		\label{fig:nesting}}
\end{figure}

The crystal structure of pristine {\dt} is shown in Fig.~\ref{fig:nesting}(a), where the buckled DyTe slabs are 
separated by the planar Te square lattice 
that tends to undergo a supercell modulation whose
periodicity interestingly depends on the Te vacancy 
concentration.\cite{Falson2024,Poddig2021}
The nesting function, $N(\bm{q})$, of the Fermi surface 
for an electron doping corresponding to the Te deficiency 
$\delta=0.2$, reported in our recent calculations
[Fig. S7(h) in Supporting information of Ref.~\cite{Falson2024}],
is replotted in Fig.~\ref{fig:nesting}(b).
It is zoomed-in near the $\bm{q}$ point corresponding to 
the {\rfive} modulation, marked as a cross point at $(0.2,0.4)$.
Two strong peaks near the cross point, 
illustrated as dark orange colors, are traced as a function
of doping which are displayed with the blue and green dots,
where their separation increases with electron doping.
The green peak is found to pass the cross point at higher
Te deficient level ($\delta>0.2$)
that does not necessarily imply the need for additional Te 
vacancies, 
since the nesting function calculations involve
several assumptions such as the rigid band approximation,
constant matrix elements,
and the exclusion of strain effect on the chemical potential.
Our calculations demonstrate that 
the nesting function peaks associated with the formation of
various supercell modulations change with chemical doping.
Especially, the nesting function peaks (denoted by blue and
green dots) approach the {\rfive} modulation $\bm{q}$ vectors
with increasing Te deficiency in a good agreement with
experiments.
%

\begin{figure} [t]
	\centering
	\includegraphics[width=8.6cm]{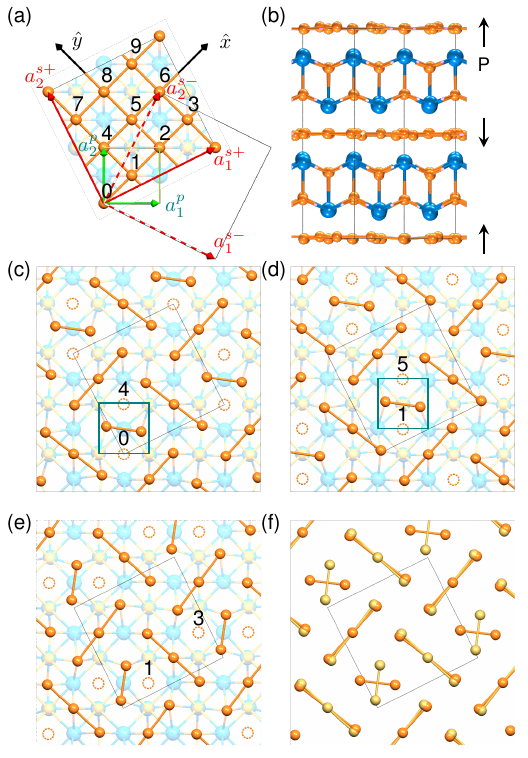}
	\caption{Structural modulation of DyTe$_2$.
		(a) Projection of the Te-square network on the (001) plane, where the numerical indices of the \rfive -modulated cell denote 
		the plausible positions of the Te$_s$ divacancies. 
		The $\bm{a}^{p}_{i}$ are the primitive lattice vectors and $\bm{a}^{s\pm}_{i}$ are two sets of supercell lattice vectors 
		with $\pm$ chirality.
		(b) Side view of $\sqrt{5}\times\sqrt{5}\times2$ modulated
		DyTe$_{1.8}$. Arrows depict the polarization configuration
		of individual Te square lattice layers in the ground state.
		(c-e) Reconstructed ordered Te-square network of DyTe$_{1.8}$ 
		for several Te divacancy positions and orientations 
		at the (c) \{0,4\}, (d) \{1,5\}, and (e) \{1,3\} sites. Small squares in (c) and (d) 
		denote the pristine cell enclosing the Te-dimer.
		(f) Mixed phase of the reconstructed Te-square network comprised of 
		the structural motifs of (c) and (e), respectively, where the large square denotes the \rfive \, cell.
		\label{fig:structure}}
\end{figure}

Consequently, in this work we focus on the {\rfive} modulation that commonly occurs
for various rare-earth elements with two Te$_s$ vacancies
per modulated cell ($\delta=0.2$)\cite{Falson2024}.
The in-plane lattice vectors of the supercell exhibiting chirality, shown in Fig.~\ref{fig:structure}(a), are
\begin{equation}
	\left(
	\begin{aligned}
		\bm{a}^{p}_{1} &=& \tfrac{a_0}{\sqrt{2}} (\hat{x}-\hat{y}) \\
		\bm{a}^{p}_{2} &=& \tfrac{a_0}{\sqrt{2}} (\hat{x}+\hat{y}) \\
	\end{aligned}
	\right.
	\qquad
	\left(
	\begin{aligned}
		\bm{a}^{s\pm}_{1} &=& 2\bm{a}^{p}_{1} \pm \bm{a}^{p}_{2} \\
		\bm{a}^{s\pm}_{2} &=& \mp\bm{a}^{p}_{1} + 2\bm{a}^{p}_{2},	
	\end{aligned}
	\right.
\end{equation}
where $\bm{a}^{p}_{i}$ are the primitive lattice vectors and $\bm{a}^{s\pm}_{i}$ are two sets of supercell lattice vectors 
with $\pm$ chirality.
In general, a synthesized sample may have both types of domains
leading to domain walls across which the chirality gets
reversed.
The two supercell structures with different chirality and 
their physical properties can be transformed to each other 
by a mirror operator, 
$\mathcal{M}_y : (x,y,z) \rightarrow (x,-y,z)$.
Unless otherwise stated throughout the remainder of this manuscript we 
focus only the $+$ chirality structure and ignore for convenience 
the superscript $^{s+}$, namely, $\bm{a}_i \equiv \bm{a}_i^{s+}$.

The numerical indices in Fig.~\ref{fig:structure}(a) represent 
the ten Te$_s$ sites of the {\rfive} supercell where the divacancy may be placed. 
The SNN divacancy (such as the $\{0,4\}$, 
denoted as the A-C configuration in Ref. \cite{Falson2024}) was 
found to have the lowest formation energy for 
the {\rfive} modulation.
The reconstructed square lattice consists of one Te-dimer and two 
Te-trimers per unit cell, resulting in 
several iso-energetic and ordered structural motifs.
Figures~\ref{fig:structure}(c) and (d) show the first and second 
structural motifs, associated with the $\{0,4\}$ and $\{1,5\}$ SNN divacancies, respectively, that look almost identical 
except for their positions relative to the DyTe slab.
The thick squares in each panel emphasize the position of the Te$_s$ dimers 
whose centers lie atop of the Dy (Te$_c$) 
atoms for the first (second) structural motif.
The two structural motifs transform into each other through the 
mirror operation, $\mathcal{M}_z$,
followed by a proper in-plane translation indicating a finite
vertical polarization (see Fig.~\ref{fig:pol}) whose sign depends 
on the position of the Te dimer, as will be discussed in the next section.
The divacancy also reduces the $\mathcal{C}_4$ rotation symmetry
of the pristine supercell to the $\mathcal{C}_2$ symmetry, 
whose rotational axis can be placed at the center of the Te dimer.
The third structural motif associated with the $\{1,3\}$ divacancy
can be transformed into the first motif of the $\{0,4\}$ divacancy through a $\mathcal{C}_4$ rotation, as illustrated in Figs.~\ref{fig:structure}(c) and (e).
%
The actual DyTe$_{1.8}$ samples may consist of a mixed phase 
of the $\{0,4\}$- and $\{1,3\}$-based structural motifs as shown 
in Fig.~\ref{fig:structure}(f).
The mixed phase can then preserve the tetragonal symmetry
in macroscopic scale while the microscopic local domains 
are under uniaxial in-plane strain 
depending on the Te dimer orientation.
This in turn raises an intriguing question whether an in-plane strain 
can align the orientation of the Te dimers, which is discussed in detail in Sec.~\ref{sec:1D}.
In general, CDWs with a commensurate cell
modulation of $n \times m$ can introduce $nm$ distinct types
of domains, associated with the relative translations 
of the modulated supercell.
Therefore, there can be five different types of domains 
for the {\rfive} modulation of DyTe$_{1.8}$.
It is noteworthy that the three structural motifs discussed
above do not transform into one another via lateral translation
alone.
The total number of distinct domains amounts to 
$2\times 2\times 5=20$,
where the first (second) factor of 2 arises from the two 
possible positions (orientations) of the Te dimer
in addition to the factor of 5 due to the 
lateral translations.


\subsection {Polarization} \label{sec:pol}

\begin{figure} [t]
	\centering
	\includegraphics[width=8.6cm]{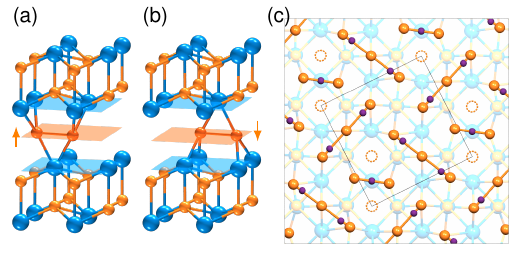}
	\caption{Polarization of the Te square network. (a) Center of Te-dimer is atop of the Dy atom corresponding to 
	the \{0,4\} divacancy configuration. (b) Center of Te-dimer is atop of the Te atom corresponding to 
	the \{1,5\} divacancy configuration.  
	The interlayer distances of the corrugated Dy-Te layers sandwiching the Te-dimer sheet are  different. 
	The Te-trimers are omitted for clarity.
	(c) Wannier charge centers lying on the covalent bonds, denoted by the purple dots on the
	reconstructed Te-square lattice for the \{0,4\} divacancy
	configuration. The position of Wannier charge centers affects the polarization as well as the dipole moments.
	\label{fig:pol}}
\end{figure}

\begin{table} [t]
	\caption{
		Dipole moment and polarization of DyTe$_{1.8}$ for the  \{0,4\} divacancy along
		the supercell lattice vectors, $\bm{a}_i$, of the
		$\sqrt{5} \times \sqrt{5} \times 1$ cell.
		\label{tab:pol}}
		\begin{ruledtabular}
		\begin{tabular}{l|cc}
		& Dipole moment ($Q_e \bm{a}_i$) & Polarization (C/m$^2$) \\
		\hline \\[-0.2cm]
		$\bm{a}_1$ &  0.500 & 0.095 \\
		$\bm{a}_2$ &  0.500 & 0.095 \\
		$\bm{a}_3$ &  0.893 & 0.160 \\
		\end{tabular}
	\end{ruledtabular}
\end{table}

The emergence of the ($\sqrt{5}\times\sqrt{5}$) superlattice induced by the periodic Te-vacant sites renders the DyTe$_{1.8}$ insulator.\cite{Falson2024} 
The formation of the divacancy breaks the inversion symmetry which in turn results in  finite polarization.
Figures~\ref{fig:pol}(a) and (b) clearly show the Dy-Te bonds of the Te$_s$ dimers for the
first and second structural motifs, which depend on their lateral position. 
Namely, in the first motif, where the center of the Te dimer is atop 
of the Dy atom, the dimer forms four (two) bonds with the Dy atoms 
on the upper (lower) layer, resulting in its proximity to the upper Dy-Te corrugated layer.
Conversely, the second structural motif exhibits the opposite behavior.

The polarization $\bm{p}$ and dipole moment $\bm{d}$
of an insulator can be computed by considering
the position of charged ions and Wannier charge centers (WCC) 
of the occupied Bloch states below the insulating 
band gap.\cite{Marzari2012,JK2019}
As described in the methodology section, the topmost 51 valence bands are Wannierized which are well separated from both
the semi-core states and the conduction bands.
The Wannier functions are found exponentially localized near the initial position of projectors.
Three and two Wannier functions are thus attached to Te$_c$ and
Te$_s$ atoms, respectively.
The positions of five interstitial Wannier functions,  
illustrated in Fig.~\ref{fig:pol}(c), lie on the
Te$_s$-Te$_s$ bonds.
The Wannier functions that deviate from the atomic sites
render the system an obstructed atomic insulator.\cite{Po_PRL2018,Radha_PRB2021,Li_PRB2022,Xu_PRB2024}
The polarization and dipole moment are given by, 
\begin{eqnarray}
	\bm{d} &=& \bm{p}V \\
	\label{eq:pol}
	&=& Q_e \left( 
	\sum_n^{\mathrm{ions}} Z_n \left( \bm{R}_n - \bm{r}_0 \right) -
	\sum_n^{\mathrm{WCC}} D \left( \bm{r}_n - \bm{r}_0 \right) 
	\right) \\
	&=& Q_e \left( 
	\sum_n^{\mathrm{ions}} Z_n \bm{R}_n -
	\sum_n^{\mathrm{WCC}} D \bm{r}_n 
	-
	\bm{r}_0 \left\{
	N^+ - N^-
	\right\}
	\right) 
	\\ 
	&=& Q_e \left( 
	\sum_n^{\mathrm{ions}} Z_n \bm{R}_n -
	\sum_n^{\mathrm{WCC}} D \bm{r}_n 
	\right),
\end{eqnarray}
where $V$ is the unit cell volume, $Q_e$ is the electron  
charge, $Z_n$  and $\bm{R}_n$ are the ionic charge number and 
atomic position of the $n$-th atom, 
$\bm{r}_n$ is the position of the $n$th WCC, and $D=2$ is the spin degeneracy of the Wannier functions.
The origin $\bm{r}_0$ does not affect the dipole moment
as it is charge neutral, $N^+ = N^-$, where
the net ionic charge, $N^+ = \sum_n^{\mathrm{ions}} Z_n$ is 
equal to the number of WCCs, $N^- = D \sum_n^{\mathrm{WCC}}$.
Note that the WCC position, $\bm{r}_n$, can be shifted by
integer multiples of lattice vectors, 
$\bm{r}_n \rightarrow \bm{r}_n + \sum_{i} m_{i} \bm{a}_{i}$ 
without affecting the physical observables  
that is associated with the gauge freedom leading to the modulo definition of the modern theory of polarization.
\cite{King-Smith1993,Vanderbilt1993,Resta1994}
The dipole moment is thus defined under modulo $D Q_e \bm{a}_i$
as long as the time reversal symmetry is preserved on the
surface as well as in the bulk.

The calculated dipole moment and polarization of the first
structural motif of the $\{0,4\}$ divacancy, are listed in
Table~\ref{tab:pol}.
The $\mathcal{C}_2$ symmetry of the crystal structure confines the 
in-plane components of the dipole moments to be quantized\cite{Zak_PRL1989}
either to 0 or $1/2$
in units of $Q_e \bm{a}_{1,2}$. 
The calculated in-plane dipole moments values of, $1/2$, indicate
nontrivial Zak phase\cite{Zak_PRL1989} of DyTe$_{1.8}$ in the 
$\sqrt{5} \times \sqrt{5} \times 1$ supercell
that induces metallic (side) surfaces due to the emergent in-gap states.
We note that the  physical samples, however, exhibit
additional $\times 2$ structural modulation along the 
$\hat{z}$ direction\cite{Falson2024}
resulting in $\sqrt{5} \times \sqrt{5} \times 2$
supercell which cancels out the in-plane dipole moment.
The concept of quantized polarization under crystal symmetry 
was first elucidated by Zak\cite{Zak_PRL1989} 
and has recently been integrated into the framework of 
topological insulators.\cite{Murakami_PRR_2020,Vanderbilt1993,Smeu_PRB,JKim_Bi}

%
This additional modulation also controls the net dipole
moment along the $\hat{z}$ direction as each layer may have
upward or downward dipole moment depending on the position of the Te$_s$ dimers of each layer.
In our previous study,\cite{Falson2024} the total energy
of the $\sqrt{5} \times \sqrt{5} \times 2$ supercell was
calculated (see Table~\ref{tab:tenergy})
for various pairs of divacancy configurations,
$\{0,4\}-\{d,d+4\}$,  $d=\{0,1,2,3,4,5\}$
where the first divacancy nucleates at sites $\{0,4\}$ on the first 
Te$_s$ layer while the second pair nucleates on sites $\{d,d+4\}$
on the second Te$_s$ layer.
The supercell structure with $d=5$ is found to exhibit the lowest energy in which the polarization direction of the upper layer is opposite to that of the lower layer 
as illustrated in Fig.~\ref{fig:structure}(b).
The next stable structure with $d=2$ has 2.4 meV/Dy higher total
energy and the polarization directions of the two layers
are parallel leading to finite net polarization,
suggesting that an out-of-plane external 
electric field, $E \gtrsim 15$ MV/m may stabilize the 
polarized structure.
	


\begin{table}
	\caption{Calculated total energy of the 
	$\sqrt{5} \times \sqrt{5} \times 2$ supercell for the lateral position, $d$,
	of the divacancy on the second Te layer relative to the first.
	The relative polarization directions of the two layers
	are labeled as P and AP for parallel and antiparallel 
	configurations, respectively.
	\label{tab:tenergy}}
	\begin{ruledtabular}
		\begin{tabular}{c|cccccc}
			& \multicolumn{6}{c}{Displacement of 2$^{\mathrm{nd}}$ layer, d} \\
			& 0 & 1 & 2 & 3 & 4 & 5 \\
			\hline \\[-0.2cm]
			$\Delta E$ (meV/Dy) &  0.0 & -6.0 & -8.4 & -5.8 & -7.3 & -10.8 \\
			Polarization & P & AP & P & AP & P & AP \\ 
		\end{tabular}
	\end{ruledtabular}
\end{table}


\subsection{Electronic structure - Van Hove singularities} \label{sec:VHS}

\begin{figure}
	\centering
	\includegraphics[width=8.6cm]{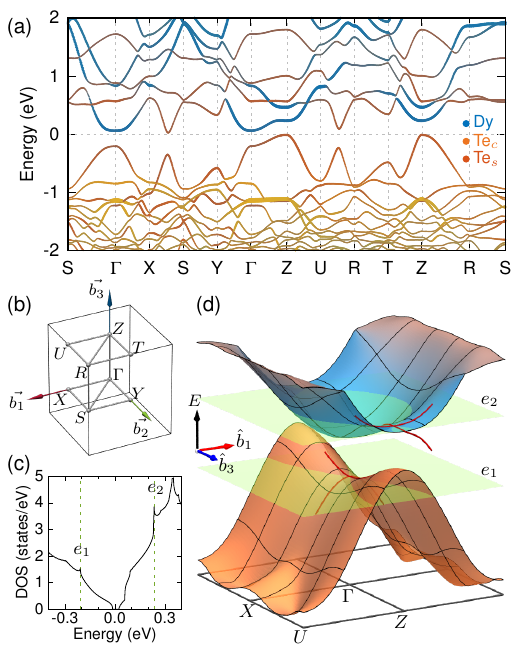}
	\caption{Calculated electronic structure of DyTe$_{1.8}$ and Van Hove singularities near the Fermi level.
	(a) Band structure of the $\{0,4\}$ divacancy configuration along high symmetry lines.
	(b) BZ of the \rfive-modulated supercell cell and the high symmetry points.
	Van Hove singularities near the Fermi level.
	(c) Calculated DOS near the Fermi level showing the emergence of Van Hove singularities at the 
	energies, $e_1$ and $e_2$, of the saddle points in 
	both the hole and electron doped regimes.
	(d) Two dimensional band structure on the $\hat{b}_1 - \hat{b}_3$ plane, where the orange and blue colors denote the Te$_s$-$p$ and Dy-$d$ orbital characters, respectively. Valence (conduction) band exhibits saddle point(s) at the $\Gamma$ (near the $Z$) point marked with red curves. 
	The energy levels, $e_1$ and $e_2$, of the saddle points are illustrated with transparent planes.
	\label{fig:VHS}}
\end{figure}

Figure~\ref{fig:VHS} shows the calculated electronic structure
of DyTe$_{1.8}$ with the $\{0,4\}$ divacancy configuration.
In Fig.~\ref{fig:VHS}(b) we plot the Brillouin zone (BZ) of the \rfive-modulated cell of DyTe$_{1.8}$ with the high symmetry directions
used in the band structure plot.
The blue, red and orange colors denote projections on the Dy-$d$, Te$_s$-$p$, and Te$_c$-p orbitals.
The indirect (direct) band gap is 0.14 (0.33) eV in qualitative agreement with the activation gap measured in transport.\cite{Falson2024} 

In our previous study\cite{Falson2024} of the 
electronic structure of pristine DyTe$_2$, 
we found that the Te$_s$-$p$ bands, originating from 
the square lattice, intersect the Fermi level and give rise to the  
nesting peaks near the $\bm{q}$ vectors corresponding 
to the \rfive\, cell modulation.
These nesting peaks explain the cell modulation and resulting
insulating band gap in DyTe$_{1.8}$, which is further
enhanced by the formation of the vacancy network, which increases 
the inter-nested-states coupling strengths.
The topmost valence band of DyTe$_{1.8}$ retains
the 2D nature of the square lattice, with highly
dispersive bands along the in-plane direction and
weakly dispersive bands along the out-of-plane direction
such as the $\Gamma-Z$ and $R-S$ symmetry lines.
The dimensional reduction in the band structure likely introduces
saddle points, leading to Van Hove singularities (VHS).
Figure~\ref{fig:VHS}(d) illustrates the saddle points 
on the $\bm{b}_1-\bm{b}_3$ plane and the corresponding 
divergent peaks of the density of states (DOS).
The conduction band, dominated by the Dy-$d$ 
orbital character, also exhibits saddle points on this plane
which slightly deviate from the $Z$ point
by $\pm 0.1 \bm{b}_3$.
%
These VHS in both the valence and conduction bands are
expected to give rise to significant optical responses due to 
the high joint DOS, 
where the energy difference, 
$e_2 - e_1 = 0.44$ eV, corresponds to the infrared wavelength, 
$\lambda = 2.8\,\mu$m.
However, these saddle points are separated in 
momentum space, necessitating a  phonon coupling to
complement the momentum difference, $\Delta k_z \sim 0.4 |\bm{b}_3|$.
Fortunately, the system undergoes another double 
cell modulation along the $\hat{z}$ direction, which 
in turn folds the $Z$ point to the $\Gamma$ point. Thus, 
in the long-wavelength limit, optical absorption between 
the two saddle points can occur without phonon assistance.
Additionally, the vertical polarization of each square 
lattice layer induces a Rashba spin splitting in the
presence of spin-orbit coupling.
The alignment of these layers' polarizations determines
the spin texture chirality of the spin-split bands,
influencing the band alignment and consequent optical
responses. Further theoretical and 
experimental investigations are needed to fully understand 
these effects.

\subsection{Electronic structure - Fermi surface} \label{sec:FS}

\begin{figure}
	\centering
	\includegraphics[width=8.6cm]{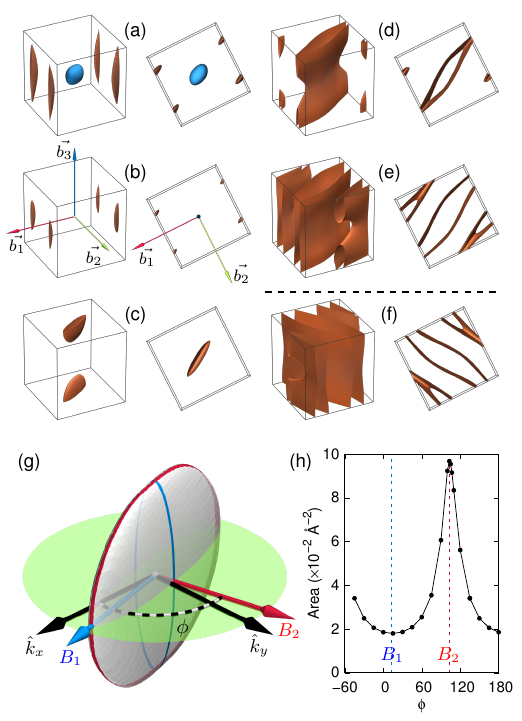}
	\caption{Asymmetric Fermi surface (FS) sheets (left columns) and their (001) projections (right columns), of  DyTe$_{1.8}$ with the  $\{0,4\}$ divacancy configuration for chemical potential shifts, 
	$\mu =E-E_F$, ranging from  (a) $+0.3$ eV to (e) $-0.5$ eV in steps of 0.2 eV.  Orange and blue colors denote the Te$_s$-$p$ and Dy-$d$ orbital characters, respectively. (f) FS sheets (left panel) and their (001) projections (right panel) for the $\{1,3\}$ divacancy configuration for $\mu=-0.5$ eV.
	(g) Fermi surface at $\mu=-0.1$ eV. Green circular area  shows the $\hat{k}_x-\hat{k}_y$ plane on which the external magnetic field is swept with an azimuthal angle $\phi$ from the $\hat{k}_x$ axis. The corresponding maximal cross-section paths under the external field along $B_1$ and $B_2$ are illustrated with blue and red stripes, respectively, on the Fermi surface.
	(h) Calculated maximal cross-section area versus $\phi$.
	\label{fig:FS}}
\end{figure}

Figures~\ref{fig:FS}(a)-(e) display the calculated 3D Fermi surface (FS) sheets and their (001) projections, of  DyTe$_{1.8}$ with the  $\{0,4\}$ divacancy configuration for chemical potential shifts, $\mu =E-E_F$, ranging from  (a) $+0.3$ eV to (e) $-0.5$ eV in steps of 0.2 eV.  The orange and blue colors denote the Te$_s$-$p$ and Dy-$d$ orbital characters, respectively. For $\mu = -0.5$ eV the chemical potential lies in the Te$_s$-$p$-derived valence bands [see Fig.~\ref{fig:VHS}(a)], and the FS consists of highly-nested parallel sheets stacked normal to the $\Gamma$-S symmetry direction of the BZ [Fig.~\ref{fig:VHS}(b)]. As the chemical potential shifts upward in energy, the FS area gradually gets reduced and isolated pockets appear at $Z$ point.
Above the Fermi level, two elongated electron pockets
emerge on the side surface of the BZ as shown in Fig.~\ref{fig:FS}(b) and eventually
a Dy-$d$ derived electron pocket emerges at $\Gamma$ point
[Fig.~\ref{fig:FS}(a)].
In Fig.~\ref{fig:FS}(f) we display the FS sheets and their (001) projections for the $\{1,3\}$ divacancy configuration for $\mu = -0.5$ eV. Comparison of Figs.~\ref{fig:FS}(e) and (f) clearly shows the 90$^{\circ}$ rotation of the FS 
associated with the Te$_s$ dimer rotation by 90$^{\circ}$ about [001], shown in Fig.~\ref{fig:structure}(e). 
Since the group velocity is always normal to the FS,
the parallel FS sheets indicate 1D conducting channels where their
direction can be switched by the Te dimer orientations, that will be
discussed in the next section.

%
As alluded above, the valence bands are dominated by the Te$_s$ atoms of the 2D square lattice.
The Te-dimer orientations switch the hopping directions within
the square lattice that further refine the 2D like band structure
into an effective 1D channel resulting in asymmetric Fermi surface
as shown in Fig.~\ref{fig:FS}.
The FS asymmetry can be confirmed by measuring the
quantum oscillations under external magnetic field whose
frequency provides information of the maximal cross-section area of the FS.
Figure~\ref{fig:FS}(g) shows the disk shape Fermi surface 
centered at the $Z$ point under hole doping for $\mu = -0.1$ eV.
The maximal cross-section area shown in Fig.~\ref{fig:FS}(h) is expected to strongly
depend on the magnetic field direction, where a peak appears at the 
azimuthal angle (measured from $\hat{k}_x$), $\phi \sim 103^{\circ}$, corresponding to $B_2$.
The red stripe on the Fermi surface shows the cyclotron
motion path under the external field $B_2$ while the
blue stripe is under $B_1$ whose cross-section area is
smaller than that of $B_2$ by a factor of five.

\subsection{Quasi 1D conducting channel} \label{sec:1D}

\begin{figure}
	\centering
	\includegraphics[width=8.6cm]{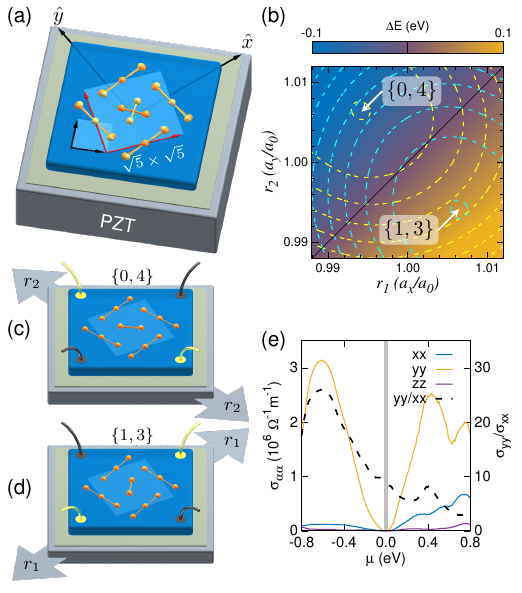}
	\caption{Effect of uniaxial in-plane strain on Te-dimer orientation. 
	(a) Schematic figure of DyTe$_{1.8}$ epitaxially grown on a piezoelectric substrate. 
	Small (large) square shows the in-plane unit cell of primitive DyTe$_2$ (\rfive-modulated DyTe$_{1.8}$) structure.
	(b) Landscape of total energy difference, $\Delta E = E(\{0,4\}) - E(\{1,3\})$,
	between the $\{0,4\}$- and $\{1,3\}$ divacancy structural motifs of the \rfive\, supercell on  two-dimensional epitaxial strain, $r_x \equiv \frac{\alpha_x}{\alpha_0}$ and 
	$r_y \equiv \frac{\alpha_y}{\alpha_0}$. Yellow and blue dashed curves indicate 
	energy contours of the $\{0,4\}$ and $\{1,3\}$ phases, respectively.
	(c) and (d) Relatively stable Te-dimer orientations under uniaxial strain. The $\{0,4\}$ ($\{1,3\}$) phase is preferred under uniaxial strain along the $\hat{y}$ ($\hat{x}$) direction. Gold (black) leads indicate higher (lower) conductivity channel.
	(e) Calculated conductivity of DyTe$_{1.8}$ with the two vacancies at $\{0,4\}$ versus the chemical potential $\mu$. Conductivity along $\hat{y}$ [see (a) or Fig.~\ref{fig:structure}(a) for the direction] is higher more than one order of magnitude compared to the other directions under hole doping. The dashed line denotes the in-plane conductivity ratio, $\sigma_{yy} / \sigma_{xx}$. The band gap near $\mu=0$ is marked with the gray box.
	\label{fig:1D}}
\end{figure}

The results of the various reconstructed iso-energetic divacancy configurations invite the 
intriguing question whether epitaxial strain, which is ubiquitous in DyTe$_{1.8}$ samples 
grown on atomically flat piezoelectric substrates, shown schematically in Fig.~\ref{fig:1D}(a),  
can stabilize a specific ordered structural motif, and whether the divacancy configuration network can 
undergo a transition from one motif to another. 
Fig.~\ref{fig:1D}(b) shows the landscape of the total energy difference,  $\Delta E = E(\{0,4\}) - E(\{1,3\})$,
between the $\{0,4\}$- and $\{1,3\}$ divacancy structural motifs of the {\rfive} reconstructed DyTe$_{1.8}$
on the two-dimensional epitaxial strain, $r_x \equiv \frac{\alpha_x}{\alpha_0}$ and 
$r_y \equiv \frac{\alpha_y}{\alpha_0}$. Here, $\alpha_0$ is the in-plane equilibrium lattice constant of the square lattice, 
and the x and y refer to the [110] and [$\bar1$10] directions, respectively. The calculations reveal that tensile epitaxial strain, $\epsilon_{yy} \approx$ 0.7\%, renders the  $\{0,4\}$ divacancy-based configuration as the ground state [Fig.~\ref{fig:1D}(c)]. On the other hand, the $\{1,3\}$ divacancy-based structural motif becomes the ground state under tensile strain, $\epsilon_{xx} \approx$ 0.7\% along [110] [Fig.~\ref{fig:1D}(d)].
Interestingly, 
these two distinct Te$_s$-vacancy ordered networks
have substantial different transport properties.
Figure~\ref{fig:1D}(e) shows 
the matrix elements of the longitudinal conductivity, $\sigma_{ii}$ ($i$=x,y,z), of DyTe$_{1.8}$ with the $\{0,4\}$ divacancy 
configuration versus the chemical potential, $\mu$. 
The conductivity ratio, 
$\frac{\sigma_{yy}}{\sigma_{xx}}$ versus $\mu$
is also plotted as a dashed line
whose scale appears on the right vertical axis.
Interestingly, $\sigma_{yy}$, is higher by more than one order of magnitude compared to $\sigma_{xx}$ 
under hole doping.

This notable conductivity switching between the two phases
upon the Te dimer orientation, suggests potential application
as a memory device.
Control of the dimer orientation can be achieved by applying
epitaxial strain via a piezoelectric substrate, which can in 
turn align all or most of the sample's Te dimers along the $\hat x$ or $\hat y$ directions, resulting 
in a colossal conductivity contrast, ranging from 1000\% to 2500\% 
between the ON and OFF phases.

Consequently, the Te dimer in the supercell acts as a molecular switch
not only in a macroscopic scale but also in a microscopic level by controlling the electron hopping directions
within the {\rfive} supercell.
The overall conductivity of the two-dimensional percolation network\cite{Sahimi_book,Cherkasova2010}
will be determined by the microscopic configurations of the Te dimer orientations, which can be tuned by epitaxial strain.
We acknowledge the significance of prior nanotube percolation experiments\cite{Du2005,Park2006}, particularly those where the nanotube orientation was controllable, which garnered considerable attention within the research community. Given the potential for strain-controlled dimer orientations, DyTe$_{1.8}$ emerges as a promising alternative platform for conducting percolation experiments.

\subsection {Migration Properties} \label{sec:migration}

\begin{figure*}
	\centering
	\includegraphics[width=17.6cm]{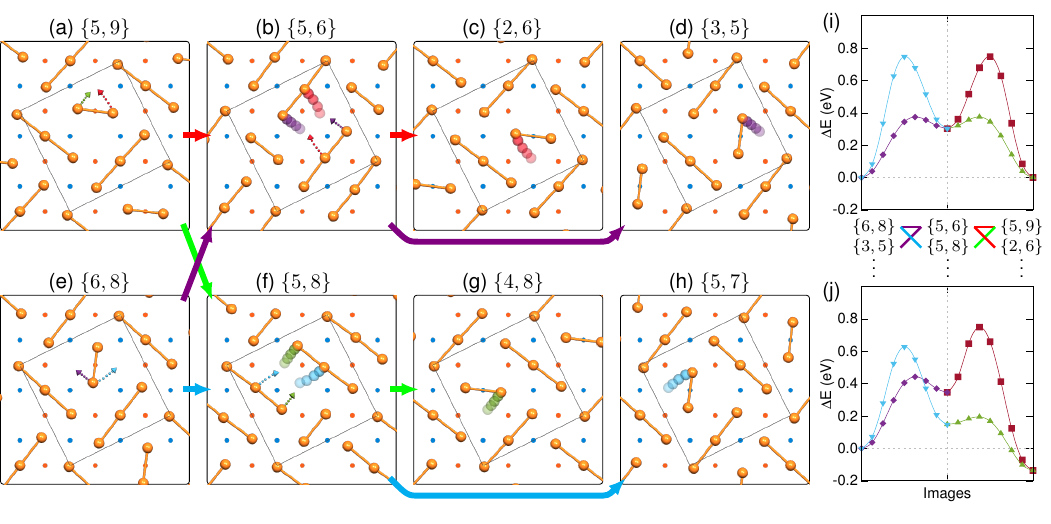}
	\caption{(a-h) Te-square net with various di-vacancy configurations and their migrations. Background blue (orange) dots denote the positions of Dy (Te$_c$) atoms below the Te-square net layer. Arrows illustrate the considered paths of ionic migrations. Four different types of migration paths are colored by red, green, cyan, and violet. (i,j) Calculated migration energy landscape of the various transitions for (i) symmetric, $r_1 = r_2 = 1.0$ and (j) asymmetric, $r_1 = 0.988$, $r_2 = 1.012$, in-plane strain.
	\label{fig:mig}}
\end{figure*}

In order to understand the diffusion properties of the divacancy configurations in the presence of absence of epitaxial strain, we have carried first principles calculations using the nudged elastic band (NEB) 
method\cite{MILLS1995305,NEB_book,Henkelman2000,Henkelman2000b}
to determine the energy barriers and the minimum energy paths associated with the atomic migrations within the Te square network. The energy saddle point identified by this method represents the barrier between two stable configurations and
the migration is assumed to occur simultaneously across all periodic (both in-plane and out-of-plane) cells, rendering the calculated energy
barrier an upper bound. Consequently, the energy required to incrementally expand a domain area could be less than the barrier determined by this method.

There are numerous atomic migration paths, some of which are symmetric equivalents with identical energy barriers. Therefore, only four irreducible migration paths require computation which collectively represent all possible migrations.
The eight dotted arrows depicted in the Fig.~\ref{fig:mig}(a), (b), (e), and (f) represent these 
four migration paths (colored red, green, cyan, and purple) 
adjacent to the vacant sites. Paths of the same color 
are reverse equivalents of each other.
A pairwise combination of these migrations can describe both polarization reversal and dimer rotation.

For instance, the migration from (a) $\{5,9\}$ to (b) $\{5,6\}$
to (c) $\{2,6\}$ changes the position of the dimer and thus
the polarization direction. In the first step, the right 
side Te of a dimer moves to the vacant site,
denoted by the red dotted arrow in Fig.~\ref{fig:mig}(a)
where the intermediate positions of the Te atom are
illustrated as transparent red spheres in Fig.~\ref{fig:mig}(b).
Next, the middle Te of a trimer moves upwards,
marked as red dotted arrow in Fig.~\ref{fig:mig}(b)
that completes the migration path to the opposite
polarization [Fig.~\ref{fig:mig}(c)].
If the second step migration is replaced by the purple
dotted arrow [Fig.~\ref{fig:mig}(b)] where the end Te atom
moves instead of the middle one of the trimer,
it arrives to the (d) $\{3,5\}$ configuration exhibiting
opposite polarization and rotated dimer orientation
compared to the initial state (a) $\{5,9\}$.

The calculated energies during each migration step
are shown in Fig.~\ref{fig:mig}(i) where the red (purple) 
colored line corresponds to the migration marked with
the red (purple) arrows with an energy barrier of
 0.75 eV (0.37 eV).
The horizontal axis is for images in the NEB method 
between stable configurations listed below the tics.
The possible migration paths between the configurations
are denoted as colored lines connecting them.
For an instance, migrations from (e) $\{6,8\}$ to
(f) $\{5,8\}$ to (a) $\{5,9\}$ are connected by cyan
and green lines below Fig.~\ref{fig:mig}(i) 
corresponding to the energy curves colored cyan and green,
respectively.

We note that the previously considered polarization
reversal migration, $\{5,9\} - \{5,6\} - \{2,6\}$ takes
the red energy curves back and forth passing 
the 0.75 eV energy barrier twice.
The Te-dimer then moves towards the $-\hat{y}$ direction.
The polarization can be reversed by another migration
of $\{5,9\} - \{5,8\} - \{4,8\}$ towards the $-\hat{x}$
direction that needs to overcome
the green energy curves passing the 0.37 eV barrier twice
instead. The latter one experiences a lower energy
barrier indicating that there is a preferred direction
in the dimer migration which is determined by the
supercell chirality and the dimer orientation.

The migration energies are considered within square symmetry, 
where pairs of paths exhibit the same energy barrier.
Figure~\ref{fig:mig} (j) shows the change of the migration energy landscape 
under uniaxial strain $r_2 > r_1$ (see the caption).
The previously degenerate total energies of the ground-state divacancy configurations are lifted, 
and the $\{5,9\}$ configuration becomes more stable than the $\{6,8\}$ due to distinct
Te-dimer orientations.
Additionally, the metastable $\{5,8\}$ configuration becomes more stable
than its $\{5,6\}$ counterpart.

\section{Summary}
The structural and electronic properties of DyTe$_{1.8}$ are investigated by employing first-principles calculations.
In addition to the {\rfive} modulation,
the detailed position and orientation of the Te dimers
are found to play a crucial role in determining the
polarization and the electronic structure near the
Fermi level.
The predicted small insulating gap with VHS in both the valence and
conduction bands imply characteristic optical responses
with fine structures related with the polarization
direction of each Te square lattice layer.
The Te-dimer-orientation-induced asymmetric Fermi surface
may be directly confirmed by quantum oscillation
measurements and paves a way to memory device
application with mechanically switchable
quasi one-dimensional conductivity.
Finally, the ionic migration paths and energy landscape results reveal a plethora of structural transitions between iso-energetic motifs.
Our predictions of these interesting and diverse properties of DyTe$_{1.8}$ will hopefully motivate further experimental and theoretical investigations.

\begin{acknowledgments}
	We appreciate insightful discussions with Joseph Falson and Adrian Llanos. 
	The work is supported by the NSF-PREP CSUN/Caltech-IQIM Partnership (grant number 2216774) and the NSF-Partnership in Research and Education in Materials (PREM) (grant number DMR-1828019). 
\end{acknowledgments}

\bibliography{DyTe_theory}

\begin{thebibliography}{43}%
\makeatletter
\providecommand \@ifxundefined [1]{%
 \@ifx{#1\undefined}
}%
\providecommand \@ifnum [1]{%
 \ifnum #1\expandafter \@firstoftwo
 \else \expandafter \@secondoftwo
 \fi
}%
\providecommand \@ifx [1]{%
 \ifx #1\expandafter \@firstoftwo
 \else \expandafter \@secondoftwo
 \fi
}%
\providecommand \natexlab [1]{#1}%
\providecommand \enquote  [1]{``#1''}%
\providecommand \bibnamefont  [1]{#1}%
\providecommand \bibfnamefont [1]{#1}%
\providecommand \citenamefont [1]{#1}%
\providecommand \href@noop [0]{\@secondoftwo}%
\providecommand \href [0]{\begingroup \@sanitize@url \@href}%
\providecommand \@href[1]{\@@startlink{#1}\@@href}%
\providecommand \@@href[1]{\endgroup#1\@@endlink}%
\providecommand \@sanitize@url [0]{\catcode `\\12\catcode `\$12\catcode
  `\&12\catcode `\#12\catcode `\^12\catcode `\_12\catcode `\%12\relax}%
\providecommand \@@startlink[1]{}%
\providecommand \@@endlink[0]{}%
\providecommand \url  [0]{\begingroup\@sanitize@url \@url }%
\providecommand \@url [1]{\endgroup\@href {#1}{\urlprefix }}%
\providecommand \urlprefix  [0]{URL }%
\providecommand \Eprint [0]{\href }%
\providecommand \doibase [0]{https://doi.org/}%
\providecommand \selectlanguage [0]{\@gobble}%
\providecommand \bibinfo  [0]{\@secondoftwo}%
\providecommand \bibfield  [0]{\@secondoftwo}%
\providecommand \translation [1]{[#1]}%
\providecommand \BibitemOpen [0]{}%
\providecommand \bibitemStop [0]{}%
\providecommand \bibitemNoStop [0]{.\EOS\space}%
\providecommand \EOS [0]{\spacefactor3000\relax}%
\providecommand \BibitemShut  [1]{\csname bibitem#1\endcsname}%
\let\auto@bib@innerbib\@empty
\bibitem [{\citenamefont {Shin}\ \emph {et~al.}(2005)\citenamefont {Shin},
  \citenamefont {Brouet}, \citenamefont {Ru}, \citenamefont {Shen},\ and\
  \citenamefont {Fisher}}]{Fisher2005}%
  \BibitemOpen
  \bibfield  {author} {\bibinfo {author} {\bibfnamefont {K.~Y.}\ \bibnamefont
  {Shin}}, \bibinfo {author} {\bibfnamefont {V.}~\bibnamefont {Brouet}},
  \bibinfo {author} {\bibfnamefont {N.}~\bibnamefont {Ru}}, \bibinfo {author}
  {\bibfnamefont {Z.~X.}\ \bibnamefont {Shen}},\ and\ \bibinfo {author}
  {\bibfnamefont {I.~R.}\ \bibnamefont {Fisher}},\ }\bibfield  {title}
  {\bibinfo {title} {Electronic structure and charge-density wave formation in
  $\mathrm{La}\mathrm{Te}_{1.95}$ and $\mathrm{Ce}\mathrm{Te}_{2.00}$},\ }\href
  {https://doi.org/10.1103/PhysRevB.72.085132} {\bibfield  {journal} {\bibinfo
  {journal} {Phys. Rev. B}\ }\textbf {\bibinfo {volume} {72}},\ \bibinfo
  {pages} {085132} (\bibinfo {year} {2005})}\BibitemShut {NoStop}%
\bibitem [{\citenamefont {Garcia}\ \emph {et~al.}(2007)\citenamefont {Garcia},
  \citenamefont {Gweon}, \citenamefont {Zhou}, \citenamefont {Graf},
  \citenamefont {Jozwiak}, \citenamefont {Jung}, \citenamefont {Kwon},\ and\
  \citenamefont {Lanzara}}]{lanzara2007}%
  \BibitemOpen
  \bibfield  {author} {\bibinfo {author} {\bibfnamefont {D.~R.}\ \bibnamefont
  {Garcia}}, \bibinfo {author} {\bibfnamefont {G.-H.}\ \bibnamefont {Gweon}},
  \bibinfo {author} {\bibfnamefont {S.~Y.}\ \bibnamefont {Zhou}}, \bibinfo
  {author} {\bibfnamefont {J.}~\bibnamefont {Graf}}, \bibinfo {author}
  {\bibfnamefont {C.~M.}\ \bibnamefont {Jozwiak}}, \bibinfo {author}
  {\bibfnamefont {M.~H.}\ \bibnamefont {Jung}}, \bibinfo {author}
  {\bibfnamefont {Y.~S.}\ \bibnamefont {Kwon}},\ and\ \bibinfo {author}
  {\bibfnamefont {A.}~\bibnamefont {Lanzara}},\ }\bibfield  {title} {\bibinfo
  {title} {Revealing charge density wave formation in the $\mathrm{LaTe}_{2}$
  system by angle resolved photoemission spectroscopy},\ }\href
  {https://doi.org/10.1103/PhysRevLett.98.166403} {\bibfield  {journal}
  {\bibinfo  {journal} {Phys. Rev. Lett.}\ }\textbf {\bibinfo {volume} {98}},\
  \bibinfo {pages} {166403} (\bibinfo {year} {2007})}\BibitemShut {NoStop}%
\bibitem [{\citenamefont {Shin}\ \emph {et~al.}(2000)\citenamefont {Shin},
  \citenamefont {Han}, \citenamefont {Min}, \citenamefont {Lee}, \citenamefont
  {Choi}, \citenamefont {Kim}, \citenamefont {Kim},\ and\ \citenamefont
  {Kwon}}]{Kwon2000}%
  \BibitemOpen
  \bibfield  {author} {\bibinfo {author} {\bibfnamefont {Y.}~\bibnamefont
  {Shin}}, \bibinfo {author} {\bibfnamefont {C.}~\bibnamefont {Han}}, \bibinfo
  {author} {\bibfnamefont {B.}~\bibnamefont {Min}}, \bibinfo {author}
  {\bibfnamefont {H.}~\bibnamefont {Lee}}, \bibinfo {author} {\bibfnamefont
  {C.}~\bibnamefont {Choi}}, \bibinfo {author} {\bibfnamefont {Y.}~\bibnamefont
  {Kim}}, \bibinfo {author} {\bibfnamefont {D.}~\bibnamefont {Kim}},\ and\
  \bibinfo {author} {\bibfnamefont {Y.}~\bibnamefont {Kwon}},\ }\bibfield
  {title} {\bibinfo {title} {Anisotropic magnetization in $\it{R}\mathrm{Te}_2$
  ($\it{R}$: $\mathrm{Ce}$, $\mathrm{Pr}$, $\mathrm{Gd}$ and $\mathrm{Sm}$)},\
  }\href {https://doi.org/https://doi.org/10.1016/S0921-4526(99)02283-8}
  {\bibfield  {journal} {\bibinfo  {journal} {Physica B: Condensed Matter}\
  }\textbf {\bibinfo {volume} {291}},\ \bibinfo {pages} {225} (\bibinfo {year}
  {2000})}\BibitemShut {NoStop}%
\bibitem [{\citenamefont {Jung}\ \emph {et~al.}(2003)\citenamefont {Jung},
  \citenamefont {Alsmadi}, \citenamefont {Kim}, \citenamefont {Bang},
  \citenamefont {Ahn}, \citenamefont {Umeo}, \citenamefont {Lacerda},
  \citenamefont {Nakotte}, \citenamefont {Ri},\ and\ \citenamefont
  {Takabatake}}]{Takabatake2003}%
  \BibitemOpen
  \bibfield  {author} {\bibinfo {author} {\bibfnamefont {M.~H.}\ \bibnamefont
  {Jung}}, \bibinfo {author} {\bibfnamefont {A.}~\bibnamefont {Alsmadi}},
  \bibinfo {author} {\bibfnamefont {H.~C.}\ \bibnamefont {Kim}}, \bibinfo
  {author} {\bibfnamefont {Y.}~\bibnamefont {Bang}}, \bibinfo {author}
  {\bibfnamefont {K.~H.}\ \bibnamefont {Ahn}}, \bibinfo {author} {\bibfnamefont
  {K.}~\bibnamefont {Umeo}}, \bibinfo {author} {\bibfnamefont {A.~H.}\
  \bibnamefont {Lacerda}}, \bibinfo {author} {\bibfnamefont {H.}~\bibnamefont
  {Nakotte}}, \bibinfo {author} {\bibfnamefont {H.~C.}\ \bibnamefont {Ri}},\
  and\ \bibinfo {author} {\bibfnamefont {T.}~\bibnamefont {Takabatake}},\
  }\bibfield  {title} {\bibinfo {title} {Superconductivity in magnetically
  ordered $\mathrm{CeTe}_{1.82}$},\ }\href
  {https://doi.org/10.1103/PhysRevB.67.212504} {\bibfield  {journal} {\bibinfo
  {journal} {Phys. Rev. B}\ }\textbf {\bibinfo {volume} {67}},\ \bibinfo
  {pages} {212504} (\bibinfo {year} {2003})}\BibitemShut {NoStop}%
\bibitem [{\citenamefont {Tremel}\ and\ \citenamefont
  {Hoffmann}(1987)}]{Hoffmann1987}%
  \BibitemOpen
  \bibfield  {author} {\bibinfo {author} {\bibfnamefont {W.}~\bibnamefont
  {Tremel}}\ and\ \bibinfo {author} {\bibfnamefont {R.}~\bibnamefont
  {Hoffmann}},\ }\bibfield  {title} {\bibinfo {title} {Square nets of
  main-group elements in solid-state materials},\ }\href
  {https://doi.org/10.1021/ja00235a021} {\bibfield  {journal} {\bibinfo
  {journal} {Journal of the American Chemical Society}\ }\textbf {\bibinfo
  {volume} {109}},\ \bibinfo {pages} {124} (\bibinfo {year}
  {1987})}\BibitemShut {NoStop}%
\bibitem [{\citenamefont {Kikuchi}(1998)}]{Kikuchi1998}%
  \BibitemOpen
  \bibfield  {author} {\bibinfo {author} {\bibfnamefont {A.}~\bibnamefont
  {Kikuchi}},\ }\bibfield  {title} {\bibinfo {title} {Electronic structure of
  lanthan ditellurides},\ }\href {https://doi.org/10.1143/JPSJ.67.1308}
  {\bibfield  {journal} {\bibinfo  {journal} {Journal of the Physical Society
  of Japan}\ }\textbf {\bibinfo {volume} {67}},\ \bibinfo {pages} {1308}
  (\bibinfo {year} {1998})},\ \Eprint
  {https://arxiv.org/abs/https://doi.org/10.1143/JPSJ.67.1308}
  {https://doi.org/10.1143/JPSJ.67.1308} \BibitemShut {NoStop}%
\bibitem [{\citenamefont {Llanos}\ \emph {et~al.}(2024)\citenamefont {Llanos},
  \citenamefont {Salmani-Rezaie}, \citenamefont {Kim}, \citenamefont
  {Kioussis}, \citenamefont {Muller},\ and\ \citenamefont
  {Falson}}]{Falson2024}%
  \BibitemOpen
  \bibfield  {author} {\bibinfo {author} {\bibfnamefont {A.}~\bibnamefont
  {Llanos}}, \bibinfo {author} {\bibfnamefont {S.}~\bibnamefont
  {Salmani-Rezaie}}, \bibinfo {author} {\bibfnamefont {J.}~\bibnamefont {Kim}},
  \bibinfo {author} {\bibfnamefont {N.}~\bibnamefont {Kioussis}}, \bibinfo
  {author} {\bibfnamefont {D.~A.}\ \bibnamefont {Muller}},\ and\ \bibinfo
  {author} {\bibfnamefont {J.}~\bibnamefont {Falson}},\ }\bibfield  {title}
  {\bibinfo {title} {Supercell formation in epitaxial rare-earth ditelluride
  thin films},\ }\href {https://doi.org/10.1021/acs.cgd.3c00755} {\bibfield
  {journal} {\bibinfo  {journal} {Crystal Growth {\&} Design}\ }\textbf
  {\bibinfo {volume} {24}},\ \bibinfo {pages} {115} (\bibinfo {year}
  {2024})}\BibitemShut {NoStop}%
\bibitem [{\citenamefont {Poddig}\ \emph {et~al.}(2021)\citenamefont {Poddig},
  \citenamefont {Gebauer}, \citenamefont {Finzel}, \citenamefont {St{\"o}we},\
  and\ \citenamefont {Doert}}]{Poddig2021}%
  \BibitemOpen
  \bibfield  {author} {\bibinfo {author} {\bibfnamefont {H.}~\bibnamefont
  {Poddig}}, \bibinfo {author} {\bibfnamefont {P.}~\bibnamefont {Gebauer}},
  \bibinfo {author} {\bibfnamefont {K.}~\bibnamefont {Finzel}}, \bibinfo
  {author} {\bibfnamefont {K.}~\bibnamefont {St{\"o}we}},\ and\ \bibinfo
  {author} {\bibfnamefont {T.}~\bibnamefont {Doert}},\ }\bibfield  {title}
  {\bibinfo {title} {Structural variations and bonding analysis of the
  rare-earth metal tellurides $\it{RE}\mathrm{Te}_{1.875 \pm \delta}$
  ($\it{RE}$ = $\mathrm{Ce}$, $\mathrm{Pr}$, $\mathrm{Sm}$, $\mathrm{Gd}$;
  $0.004 \le \delta \le 0.025$)},\ }\href
  {https://doi.org/10.1021/acs.inorgchem.1c01230} {\bibfield  {journal}
  {\bibinfo  {journal} {Inorganic Chemistry}\ }\textbf {\bibinfo {volume}
  {60}},\ \bibinfo {pages} {11231} (\bibinfo {year} {2021})}\BibitemShut
  {NoStop}%
\bibitem [{\citenamefont {Plambeck-Fischer}\ \emph {et~al.}(1989)\citenamefont
  {Plambeck-Fischer}, \citenamefont {Abriel},\ and\ \citenamefont
  {Urland}}]{Plambeck-Fischer1989}%
  \BibitemOpen
  \bibfield  {author} {\bibinfo {author} {\bibfnamefont {P.}~\bibnamefont
  {Plambeck-Fischer}}, \bibinfo {author} {\bibfnamefont {W.}~\bibnamefont
  {Abriel}},\ and\ \bibinfo {author} {\bibfnamefont {W.}~\bibnamefont
  {Urland}},\ }\bibfield  {title} {\bibinfo {title} {Preparation and crystal
  structure of $\it{RE}\mathrm{Se}_{1.9}$ ($\it{RE}$ = $\mathrm{Ce}$,
  $\mathrm{Pr}$)},\ }\href
  {https://doi.org/https://doi.org/10.1016/0022-4596(89)90140-0} {\bibfield
  {journal} {\bibinfo  {journal} {Journal of Solid State Chemistry}\ }\textbf
  {\bibinfo {volume} {78}},\ \bibinfo {pages} {164} (\bibinfo {year}
  {1989})}\BibitemShut {NoStop}%
\bibitem [{\citenamefont {Dashjav}\ \emph {et~al.}(2000)\citenamefont
  {Dashjav}, \citenamefont {Oeckler}, \citenamefont {Doert}, \citenamefont
  {Mattausch},\ and\ \citenamefont {Böttcher}}]{Dashav2000}%
  \BibitemOpen
  \bibfield  {author} {\bibinfo {author} {\bibfnamefont {E.}~\bibnamefont
  {Dashjav}}, \bibinfo {author} {\bibfnamefont {O.}~\bibnamefont {Oeckler}},
  \bibinfo {author} {\bibfnamefont {T.}~\bibnamefont {Doert}}, \bibinfo
  {author} {\bibfnamefont {H.}~\bibnamefont {Mattausch}},\ and\ \bibinfo
  {author} {\bibfnamefont {P.}~\bibnamefont {Böttcher}},\ }\bibfield  {title}
  {\bibinfo {title} {$\mathrm{Gd}_8\mathrm{Se}_{15}$—$\mathrm{A}$ 24-fold
  superstructure of the $\mathrm{ZrSSi}$ type},\ }\href
  {https://doi.org/https://doi.org/10.1002/1521-3773(20000602)39:11<1987::AID-ANIE1987>3.0.CO;2-4}
  {\bibfield  {journal} {\bibinfo  {journal} {Angewandte Chemie International
  Edition}\ }\textbf {\bibinfo {volume} {39}},\ \bibinfo {pages} {1987}
  (\bibinfo {year} {2000})},\ \Eprint
  {https://arxiv.org/abs/https://onlinelibrary.wiley.com/doi/pdf/10.1002/1521-3773}
  {https://onlinelibrary.wiley.com/doi/pdf/10.1002/1521-3773} \BibitemShut
  {NoStop}%
\bibitem [{\citenamefont {van~der Lee}\ \emph {et~al.}(1997)\citenamefont
  {van~der Lee}, \citenamefont {Hoistad}, \citenamefont {Evain}, \citenamefont
  {Foran},\ and\ \citenamefont {Lee}}]{DerLee1997}%
  \BibitemOpen
  \bibfield  {author} {\bibinfo {author} {\bibfnamefont {A.}~\bibnamefont
  {van~der Lee}}, \bibinfo {author} {\bibfnamefont {L.~M.}\ \bibnamefont
  {Hoistad}}, \bibinfo {author} {\bibfnamefont {M.}~\bibnamefont {Evain}},
  \bibinfo {author} {\bibfnamefont {B.~J.}\ \bibnamefont {Foran}},\ and\
  \bibinfo {author} {\bibfnamefont {S.}~\bibnamefont {Lee}},\ }\bibfield
  {title} {\bibinfo {title} {Resolution of the 66-fold superstructure of
  $\mathrm{DySe}_{1.84}$ by x-ray diffraction and second-moment scaled
  h{\"u}ckel calculations},\ }\href {https://doi.org/10.1021/cm960302t}
  {\bibfield  {journal} {\bibinfo  {journal} {Chemistry of Materials}\ }\textbf
  {\bibinfo {volume} {9}},\ \bibinfo {pages} {218} (\bibinfo {year}
  {1997})}\BibitemShut {NoStop}%
\bibitem [{\citenamefont {Graf}\ and\ \citenamefont {Doert}(2009)}]{Graf2009}%
  \BibitemOpen
  \bibfield  {author} {\bibinfo {author} {\bibfnamefont {C.}~\bibnamefont
  {Graf}}\ and\ \bibinfo {author} {\bibfnamefont {T.}~\bibnamefont {Doert}},\
  }\bibfield  {title} {\bibinfo {title} {$\mathrm{LaSe}_{1.85}$,
  $\mathrm{CeSe}_{1.83}$, $\mathrm{NdSe}_{1.83}$ and $\mathrm{SmSe}_{1.84}$ –
  four new rare earth metal polyselenides with incommensurate site occupancy
  and displacive modulation},\ }\href
  {https://doi.org/doi:10.1524/zkri.2009.1197} {\bibfield  {journal} {\bibinfo
  {journal} {Zeitschrift für Kristallographie}\ }\textbf {\bibinfo {volume}
  {224}},\ \bibinfo {pages} {568} (\bibinfo {year} {2009})}\BibitemShut
  {NoStop}%
\bibitem [{\citenamefont {Lee}\ and\ \citenamefont {Foran}(1996)}]{Lee1996}%
  \BibitemOpen
  \bibfield  {author} {\bibinfo {author} {\bibfnamefont {S.}~\bibnamefont
  {Lee}}\ and\ \bibinfo {author} {\bibfnamefont {B.~J.}\ \bibnamefont
  {Foran}},\ }\bibfield  {title} {\bibinfo {title} {Defective lattice charge
  density waves in $\mathrm{La}_{10}\mathrm{Se}_{19}$,
  $\mathrm{Cs}_3\mathrm{Te}_{22}$, $\mathrm{RbDy}_3\mathrm{Se}_8$, and
  $\mathrm{Dy}_{65.33}\mathrm{Se}_{120}$},\ }\href
  {https://doi.org/10.1021/ja951965z} {\bibfield  {journal} {\bibinfo
  {journal} {Journal of the American Chemical Society}\ }\textbf {\bibinfo
  {volume} {118}},\ \bibinfo {pages} {9139} (\bibinfo {year}
  {1996})}\BibitemShut {NoStop}%
\bibitem [{\citenamefont {Ijjaali}\ and\ \citenamefont
  {Ibers}(2006)}]{Ijjaali2006}%
  \BibitemOpen
  \bibfield  {author} {\bibinfo {author} {\bibfnamefont {I.}~\bibnamefont
  {Ijjaali}}\ and\ \bibinfo {author} {\bibfnamefont {J.~A.}\ \bibnamefont
  {Ibers}},\ }\bibfield  {title} {\bibinfo {title} {Two new binary lanthanide
  polytellurides: Syntheses and crystal structures of $\mathrm{CeTe}_{1.90}$
  and $\mathrm{SmTe}_{1.80}$},\ }\href
  {https://doi.org/https://doi.org/10.1016/j.jssc.2006.07.010} {\bibfield
  {journal} {\bibinfo  {journal} {Journal of Solid State Chemistry}\ }\textbf
  {\bibinfo {volume} {179}},\ \bibinfo {pages} {3456} (\bibinfo {year}
  {2006})}\BibitemShut {NoStop}%
\bibitem [{\citenamefont {Kresse}\ and\ \citenamefont
  {Furthm\"uller}(1996{\natexlab{a}})}]{Kresse1996}%
  \BibitemOpen
  \bibfield  {author} {\bibinfo {author} {\bibfnamefont {G.}~\bibnamefont
  {Kresse}}\ and\ \bibinfo {author} {\bibfnamefont {J.}~\bibnamefont
  {Furthm\"uller}},\ }\bibfield  {title} {\bibinfo {title} {Efficient iterative
  schemes for $\it{ab initio}$ total-energy calculations using a plane-wave
  basis set},\ }\href {https://doi.org/10.1103/PhysRevB.54.11169} {\bibfield
  {journal} {\bibinfo  {journal} {Phys. Rev. B}\ }\textbf {\bibinfo {volume}
  {54}},\ \bibinfo {pages} {11169} (\bibinfo {year}
  {1996}{\natexlab{a}})}\BibitemShut {NoStop}%
\bibitem [{\citenamefont {Kresse}\ and\ \citenamefont
  {Furthm\"uller}(1996{\natexlab{b}})}]{Kresse1996b}%
  \BibitemOpen
  \bibfield  {author} {\bibinfo {author} {\bibfnamefont {G.}~\bibnamefont
  {Kresse}}\ and\ \bibinfo {author} {\bibfnamefont {J.}~\bibnamefont
  {Furthm\"uller}},\ }\bibfield  {title} {\bibinfo {title} {Efficiency of
  ab-initio total energy calculations for metals and semiconductors using a
  plane-wave basis set},\ }\href
  {https://doi.org/https://doi.org/10.1016/0927-0256(96)00008-0} {\bibfield
  {journal} {\bibinfo  {journal} {Computational Materials Science}\ }\textbf
  {\bibinfo {volume} {6}},\ \bibinfo {pages} {15 } (\bibinfo {year}
  {1996}{\natexlab{b}})}\BibitemShut {NoStop}%
\bibitem [{\citenamefont {Bl\"ochl}(1994)}]{Blochl1994}%
  \BibitemOpen
  \bibfield  {author} {\bibinfo {author} {\bibfnamefont {P.~E.}\ \bibnamefont
  {Bl\"ochl}},\ }\bibfield  {title} {\bibinfo {title} {Projector augmented-wave
  method},\ }\href {https://doi.org/10.1103/PhysRevB.50.17953} {\bibfield
  {journal} {\bibinfo  {journal} {Phys. Rev. B}\ }\textbf {\bibinfo {volume}
  {50}},\ \bibinfo {pages} {17953} (\bibinfo {year} {1994})}\BibitemShut
  {NoStop}%
\bibitem [{\citenamefont {Kresse}\ and\ \citenamefont
  {Joubert}(1999)}]{Kresse1999}%
  \BibitemOpen
  \bibfield  {author} {\bibinfo {author} {\bibfnamefont {G.}~\bibnamefont
  {Kresse}}\ and\ \bibinfo {author} {\bibfnamefont {D.}~\bibnamefont
  {Joubert}},\ }\bibfield  {title} {\bibinfo {title} {From ultrasoft
  pseudopotentials to the projector augmented-wave method},\ }\href
  {https://doi.org/10.1103/PhysRevB.59.1758} {\bibfield  {journal} {\bibinfo
  {journal} {Phys. Rev. B}\ }\textbf {\bibinfo {volume} {59}},\ \bibinfo
  {pages} {1758} (\bibinfo {year} {1999})}\BibitemShut {NoStop}%
\bibitem [{\citenamefont {Perdew}\ \emph {et~al.}(2008)\citenamefont {Perdew},
  \citenamefont {Ruzsinszky}, \citenamefont {Csonka}, \citenamefont {Vydrov},
  \citenamefont {Scuseria}, \citenamefont {Constantin}, \citenamefont {Zhou},\
  and\ \citenamefont {Burke}}]{Perdew2008}%
  \BibitemOpen
  \bibfield  {author} {\bibinfo {author} {\bibfnamefont {J.~P.}\ \bibnamefont
  {Perdew}}, \bibinfo {author} {\bibfnamefont {A.}~\bibnamefont {Ruzsinszky}},
  \bibinfo {author} {\bibfnamefont {G.~I.}\ \bibnamefont {Csonka}}, \bibinfo
  {author} {\bibfnamefont {O.~A.}\ \bibnamefont {Vydrov}}, \bibinfo {author}
  {\bibfnamefont {G.~E.}\ \bibnamefont {Scuseria}}, \bibinfo {author}
  {\bibfnamefont {L.~A.}\ \bibnamefont {Constantin}}, \bibinfo {author}
  {\bibfnamefont {X.}~\bibnamefont {Zhou}},\ and\ \bibinfo {author}
  {\bibfnamefont {K.}~\bibnamefont {Burke}},\ }\bibfield  {title} {\bibinfo
  {title} {Restoring the density-gradient expansion for exchange in solids and
  surfaces},\ }\href {https://doi.org/10.1103/PhysRevLett.100.136406}
  {\bibfield  {journal} {\bibinfo  {journal} {Phys. Rev. Lett.}\ }\textbf
  {\bibinfo {volume} {100}},\ \bibinfo {pages} {136406} (\bibinfo {year}
  {2008})}\BibitemShut {NoStop}%
\bibitem [{\citenamefont {Mostofi}\ \emph {et~al.}(2014)\citenamefont
  {Mostofi}, \citenamefont {Yates}, \citenamefont {Pizzi}, \citenamefont {Lee},
  \citenamefont {Souza}, \citenamefont {Vanderbilt},\ and\ \citenamefont
  {Marzari}}]{Mostofi2014}%
  \BibitemOpen
  \bibfield  {author} {\bibinfo {author} {\bibfnamefont {A.~A.}\ \bibnamefont
  {Mostofi}}, \bibinfo {author} {\bibfnamefont {J.~R.}\ \bibnamefont {Yates}},
  \bibinfo {author} {\bibfnamefont {G.}~\bibnamefont {Pizzi}}, \bibinfo
  {author} {\bibfnamefont {Y.-S.}\ \bibnamefont {Lee}}, \bibinfo {author}
  {\bibfnamefont {I.}~\bibnamefont {Souza}}, \bibinfo {author} {\bibfnamefont
  {D.}~\bibnamefont {Vanderbilt}},\ and\ \bibinfo {author} {\bibfnamefont
  {N.}~\bibnamefont {Marzari}},\ }\bibfield  {title} {\bibinfo {title} {An
  updated version of wannier90: A tool for obtaining maximally-localised
  wannier functions},\ }\href
  {https://doi.org/https://doi.org/10.1016/j.cpc.2014.05.003} {\bibfield
  {journal} {\bibinfo  {journal} {Comput. Phys. Commun.}\ }\textbf {\bibinfo
  {volume} {185}},\ \bibinfo {pages} {2309 } (\bibinfo {year}
  {2014})}\BibitemShut {NoStop}%
\bibitem [{\citenamefont {Marzari}\ \emph {et~al.}(2012)\citenamefont
  {Marzari}, \citenamefont {Mostofi}, \citenamefont {Yates}, \citenamefont
  {Souza},\ and\ \citenamefont {Vanderbilt}}]{Marzari2012}%
  \BibitemOpen
  \bibfield  {author} {\bibinfo {author} {\bibfnamefont {N.}~\bibnamefont
  {Marzari}}, \bibinfo {author} {\bibfnamefont {A.~A.}\ \bibnamefont
  {Mostofi}}, \bibinfo {author} {\bibfnamefont {J.~R.}\ \bibnamefont {Yates}},
  \bibinfo {author} {\bibfnamefont {I.}~\bibnamefont {Souza}},\ and\ \bibinfo
  {author} {\bibfnamefont {D.}~\bibnamefont {Vanderbilt}},\ }\bibfield  {title}
  {\bibinfo {title} {Maximally localized {Wannier} functions: {Theory} and
  applications},\ }\href {https://doi.org/10.1103/RevModPhys.84.1419}
  {\bibfield  {journal} {\bibinfo  {journal} {Rev. Mod. Phys.}\ }\textbf
  {\bibinfo {volume} {84}},\ \bibinfo {pages} {1419} (\bibinfo {year}
  {2012})}\BibitemShut {NoStop}%
\bibitem [{\citenamefont {Becke}\ and\ \citenamefont
  {Johnson}(2006)}]{Becke2006}%
  \BibitemOpen
  \bibfield  {author} {\bibinfo {author} {\bibfnamefont {A.~D.}\ \bibnamefont
  {Becke}}\ and\ \bibinfo {author} {\bibfnamefont {E.~R.}\ \bibnamefont
  {Johnson}},\ }\bibfield  {title} {\bibinfo {title} {{A simple effective
  potential for exchange}},\ }\bibfield  {journal} {\bibinfo  {journal} {The
  Journal of Chemical Physics}\ }\textbf {\bibinfo {volume} {124}},\ \href
  {https://doi.org/10.1063/1.2213970} {10.1063/1.2213970} (\bibinfo {year}
  {2006}),\ \bibinfo {note} {221101},\ \Eprint
  {https://arxiv.org/abs/https://pubs.aip.org/aip/jcp/article-pdf/doi/10.1063/1.2213970/15385734/221101\_1\_online.pdf}
  {https://pubs.aip.org/aip/jcp/article-pdf/doi/10.1063/1.2213970/15385734/221101\_1\_online.pdf}
  \BibitemShut {NoStop}%
\bibitem [{\citenamefont {Tran}\ and\ \citenamefont {Blaha}(2009)}]{Tran2009}%
  \BibitemOpen
  \bibfield  {author} {\bibinfo {author} {\bibfnamefont {F.}~\bibnamefont
  {Tran}}\ and\ \bibinfo {author} {\bibfnamefont {P.}~\bibnamefont {Blaha}},\
  }\bibfield  {title} {\bibinfo {title} {Accurate band gaps of semiconductors
  and insulators with a semilocal exchange-correlation potential},\ }\href
  {https://doi.org/10.1103/PhysRevLett.102.226401} {\bibfield  {journal}
  {\bibinfo  {journal} {Phys. Rev. Lett.}\ }\textbf {\bibinfo {volume} {102}},\
  \bibinfo {pages} {226401} (\bibinfo {year} {2009})}\BibitemShut {NoStop}%
\bibitem [{\citenamefont {Kim}\ \emph {et~al.}(2019)\citenamefont {Kim},
  \citenamefont {Rabe},\ and\ \citenamefont {Vanderbilt}}]{JK2019}%
  \BibitemOpen
  \bibfield  {author} {\bibinfo {author} {\bibfnamefont {J.}~\bibnamefont
  {Kim}}, \bibinfo {author} {\bibfnamefont {K.~M.}\ \bibnamefont {Rabe}},\ and\
  \bibinfo {author} {\bibfnamefont {D.}~\bibnamefont {Vanderbilt}},\ }\bibfield
   {title} {\bibinfo {title} {Negative piezoelectric response of van der waals
  layered bismuth tellurohalides},\ }\href
  {https://doi.org/10.1103/PhysRevB.100.104115} {\bibfield  {journal} {\bibinfo
   {journal} {Phys. Rev. B}\ }\textbf {\bibinfo {volume} {100}},\ \bibinfo
  {pages} {104115} (\bibinfo {year} {2019})}\BibitemShut {NoStop}%
\bibitem [{\citenamefont {Po}\ \emph {et~al.}(2018)\citenamefont {Po},
  \citenamefont {Watanabe},\ and\ \citenamefont {Vishwanath}}]{Po_PRL2018}%
  \BibitemOpen
  \bibfield  {author} {\bibinfo {author} {\bibfnamefont {H.~C.}\ \bibnamefont
  {Po}}, \bibinfo {author} {\bibfnamefont {H.}~\bibnamefont {Watanabe}},\ and\
  \bibinfo {author} {\bibfnamefont {A.}~\bibnamefont {Vishwanath}},\ }\bibfield
   {title} {\bibinfo {title} {Fragile topology and wannier obstructions},\
  }\href {https://doi.org/10.1103/PhysRevLett.121.126402} {\bibfield  {journal}
  {\bibinfo  {journal} {Phys. Rev. Lett.}\ }\textbf {\bibinfo {volume} {121}},\
  \bibinfo {pages} {126402} (\bibinfo {year} {2018})}\BibitemShut {NoStop}%
\bibitem [{\citenamefont {Radha}\ and\ \citenamefont
  {Lambrecht}(2021)}]{Radha_PRB2021}%
  \BibitemOpen
  \bibfield  {author} {\bibinfo {author} {\bibfnamefont {S.~K.}\ \bibnamefont
  {Radha}}\ and\ \bibinfo {author} {\bibfnamefont {W.~R.~L.}\ \bibnamefont
  {Lambrecht}},\ }\bibfield  {title} {\bibinfo {title} {Topological obstructed
  atomic limit insulators by annihilating dirac fermions},\ }\href
  {https://doi.org/10.1103/PhysRevB.103.075435} {\bibfield  {journal} {\bibinfo
   {journal} {Phys. Rev. B}\ }\textbf {\bibinfo {volume} {103}},\ \bibinfo
  {pages} {075435} (\bibinfo {year} {2021})}\BibitemShut {NoStop}%
\bibitem [{\citenamefont {Li}\ \emph {et~al.}(2022)\citenamefont {Li},
  \citenamefont {Ma}, \citenamefont {Liu}, \citenamefont {Yu},\ and\
  \citenamefont {Yao}}]{Li_PRB2022}%
  \BibitemOpen
  \bibfield  {author} {\bibinfo {author} {\bibfnamefont {X.-P.}\ \bibnamefont
  {Li}}, \bibinfo {author} {\bibfnamefont {D.-S.}\ \bibnamefont {Ma}}, \bibinfo
  {author} {\bibfnamefont {C.-C.}\ \bibnamefont {Liu}}, \bibinfo {author}
  {\bibfnamefont {Z.-M.}\ \bibnamefont {Yu}},\ and\ \bibinfo {author}
  {\bibfnamefont {Y.}~\bibnamefont {Yao}},\ }\bibfield  {title} {\bibinfo
  {title} {From atomic semimetal to topological nontrivial insulator},\ }\href
  {https://doi.org/10.1103/PhysRevB.105.165135} {\bibfield  {journal} {\bibinfo
   {journal} {Phys. Rev. B}\ }\textbf {\bibinfo {volume} {105}},\ \bibinfo
  {pages} {165135} (\bibinfo {year} {2022})}\BibitemShut {NoStop}%
\bibitem [{\citenamefont {Xu}\ \emph {et~al.}(2024)\citenamefont {Xu},
  \citenamefont {Elcoro}, \citenamefont {Song}, \citenamefont {Vergniory},
  \citenamefont {Felser}, \citenamefont {Parkin}, \citenamefont {Regnault},
  \citenamefont {Ma\~nes},\ and\ \citenamefont {Bernevig}}]{Xu_PRB2024}%
  \BibitemOpen
  \bibfield  {author} {\bibinfo {author} {\bibfnamefont {Y.}~\bibnamefont
  {Xu}}, \bibinfo {author} {\bibfnamefont {L.}~\bibnamefont {Elcoro}}, \bibinfo
  {author} {\bibfnamefont {Z.-D.}\ \bibnamefont {Song}}, \bibinfo {author}
  {\bibfnamefont {M.~G.}\ \bibnamefont {Vergniory}}, \bibinfo {author}
  {\bibfnamefont {C.}~\bibnamefont {Felser}}, \bibinfo {author} {\bibfnamefont
  {S.~S.~P.}\ \bibnamefont {Parkin}}, \bibinfo {author} {\bibfnamefont
  {N.}~\bibnamefont {Regnault}}, \bibinfo {author} {\bibfnamefont {J.~L.}\
  \bibnamefont {Ma\~nes}},\ and\ \bibinfo {author} {\bibfnamefont {B.~A.}\
  \bibnamefont {Bernevig}},\ }\bibfield  {title} {\bibinfo {title}
  {Filling-enforced obstructed atomic insulators},\ }\href
  {https://doi.org/10.1103/PhysRevB.109.165139} {\bibfield  {journal} {\bibinfo
   {journal} {Phys. Rev. B}\ }\textbf {\bibinfo {volume} {109}},\ \bibinfo
  {pages} {165139} (\bibinfo {year} {2024})}\BibitemShut {NoStop}%
\bibitem [{\citenamefont {King-Smith}\ and\ \citenamefont
  {Vanderbilt}(1993)}]{King-Smith1993}%
  \BibitemOpen
  \bibfield  {author} {\bibinfo {author} {\bibfnamefont {R.~D.}\ \bibnamefont
  {King-Smith}}\ and\ \bibinfo {author} {\bibfnamefont {D.}~\bibnamefont
  {Vanderbilt}},\ }\bibfield  {title} {\bibinfo {title} {Theory of polarization
  of crystalline solids},\ }\href {https://doi.org/10.1103/PhysRevB.47.1651}
  {\bibfield  {journal} {\bibinfo  {journal} {Phys. Rev. B}\ }\textbf {\bibinfo
  {volume} {47}},\ \bibinfo {pages} {1651} (\bibinfo {year}
  {1993})}\BibitemShut {NoStop}%
\bibitem [{\citenamefont {Vanderbilt}\ and\ \citenamefont
  {King-Smith}(1993)}]{Vanderbilt1993}%
  \BibitemOpen
  \bibfield  {author} {\bibinfo {author} {\bibfnamefont {D.}~\bibnamefont
  {Vanderbilt}}\ and\ \bibinfo {author} {\bibfnamefont {R.~D.}\ \bibnamefont
  {King-Smith}},\ }\bibfield  {title} {\bibinfo {title} {Electric polarization
  as a bulk quantity and its relation to surface charge},\ }\href
  {https://doi.org/10.1103/PhysRevB.48.4442} {\bibfield  {journal} {\bibinfo
  {journal} {Phys. Rev. B}\ }\textbf {\bibinfo {volume} {48}},\ \bibinfo
  {pages} {4442} (\bibinfo {year} {1993})}\BibitemShut {NoStop}%
\bibitem [{\citenamefont {Resta}(1994)}]{Resta1994}%
  \BibitemOpen
  \bibfield  {author} {\bibinfo {author} {\bibfnamefont {R.}~\bibnamefont
  {Resta}},\ }\bibfield  {title} {\bibinfo {title} {Macroscopic polarization in
  crystalline dielectrics: the geometric phase approach},\ }\href
  {https://doi.org/10.1103/RevModPhys.66.899} {\bibfield  {journal} {\bibinfo
  {journal} {Rev. Mod. Phys.}\ }\textbf {\bibinfo {volume} {66}},\ \bibinfo
  {pages} {899} (\bibinfo {year} {1994})}\BibitemShut {NoStop}%
\bibitem [{\citenamefont {Zak}(1989)}]{Zak_PRL1989}%
  \BibitemOpen
  \bibfield  {author} {\bibinfo {author} {\bibfnamefont {J.}~\bibnamefont
  {Zak}},\ }\bibfield  {title} {\bibinfo {title} {Berry's phase for energy
  bands in solids},\ }\href {https://doi.org/10.1103/PhysRevLett.62.2747}
  {\bibfield  {journal} {\bibinfo  {journal} {Phys. Rev. Lett.}\ }\textbf
  {\bibinfo {volume} {62}},\ \bibinfo {pages} {2747} (\bibinfo {year}
  {1989})}\BibitemShut {NoStop}%
\bibitem [{\citenamefont {Aihara}\ \emph {et~al.}(2020)\citenamefont {Aihara},
  \citenamefont {Hirayama},\ and\ \citenamefont
  {Murakami}}]{Murakami_PRR_2020}%
  \BibitemOpen
  \bibfield  {author} {\bibinfo {author} {\bibfnamefont {Y.}~\bibnamefont
  {Aihara}}, \bibinfo {author} {\bibfnamefont {M.}~\bibnamefont {Hirayama}},\
  and\ \bibinfo {author} {\bibfnamefont {S.}~\bibnamefont {Murakami}},\
  }\bibfield  {title} {\bibinfo {title} {Anomalous dielectric response in
  insulators with the $\ensuremath{\pi}$ zak phase},\ }\href
  {https://doi.org/10.1103/PhysRevResearch.2.033224} {\bibfield  {journal}
  {\bibinfo  {journal} {Phys. Rev. Research}\ }\textbf {\bibinfo {volume}
  {2}},\ \bibinfo {pages} {033224} (\bibinfo {year} {2020})}\BibitemShut
  {NoStop}%
\bibitem [{\citenamefont {Smeu}\ \emph {et~al.}(2012)\citenamefont {Smeu},
  \citenamefont {Guo}, \citenamefont {Ji},\ and\ \citenamefont
  {Wolkow}}]{Smeu_PRB}%
  \BibitemOpen
  \bibfield  {author} {\bibinfo {author} {\bibfnamefont {M.}~\bibnamefont
  {Smeu}}, \bibinfo {author} {\bibfnamefont {H.}~\bibnamefont {Guo}}, \bibinfo
  {author} {\bibfnamefont {W.}~\bibnamefont {Ji}},\ and\ \bibinfo {author}
  {\bibfnamefont {R.~A.}\ \bibnamefont {Wolkow}},\ }\bibfield  {title}
  {\bibinfo {title} {Electronic properties of si(111)-$7 \times 7$ and related
  reconstructions: Density functional theory calculations},\ }\href
  {https://doi.org/10.1103/PhysRevB.85.195315} {\bibfield  {journal} {\bibinfo
  {journal} {Phys. Rev. B}\ }\textbf {\bibinfo {volume} {85}},\ \bibinfo
  {pages} {195315} (\bibinfo {year} {2012})}\BibitemShut {NoStop}%
\bibitem [{\citenamefont {Kim}\ \emph {et~al.}(2023)\citenamefont {Kim},
  \citenamefont {Huang}, \citenamefont {Lin}, \citenamefont {Vanderbilt},\ and\
  \citenamefont {Kioussis}}]{JKim_Bi}%
  \BibitemOpen
  \bibfield  {author} {\bibinfo {author} {\bibfnamefont {J.}~\bibnamefont
  {Kim}}, \bibinfo {author} {\bibfnamefont {C.-Y.}\ \bibnamefont {Huang}},
  \bibinfo {author} {\bibfnamefont {H.}~\bibnamefont {Lin}}, \bibinfo {author}
  {\bibfnamefont {D.}~\bibnamefont {Vanderbilt}},\ and\ \bibinfo {author}
  {\bibfnamefont {N.}~\bibnamefont {Kioussis}},\ }\bibfield  {title} {\bibinfo
  {title} {Bismuth antiphase domain wall: A three-dimensional manifestation of
  the su-schrieffer-heeger model},\ }\href
  {https://doi.org/10.1103/PhysRevB.107.045135} {\bibfield  {journal} {\bibinfo
   {journal} {Phys. Rev. B}\ }\textbf {\bibinfo {volume} {107}},\ \bibinfo
  {pages} {045135} (\bibinfo {year} {2023})}\BibitemShut {NoStop}%
\bibitem [{\citenamefont {Sahimi}(2023)}]{Sahimi_book}%
  \BibitemOpen
  \bibfield  {author} {\bibinfo {author} {\bibfnamefont {M.}~\bibnamefont
  {Sahimi}},\ }\href
  {https://doi.org/https://doi.org/10.1007/978-3-031-20386-2} {\emph {\bibinfo
  {title} {Application of Percolation Theory}}},\ Vol.\ \bibinfo {volume}
  {213}\ (\bibinfo  {publisher} {SpringerCham},\ \bibinfo {year} {2023})\ pp.\
  \bibinfo {pages} {XXI,680}\BibitemShut {NoStop}%
\bibitem [{\citenamefont {Cherkasova}\ \emph {et~al.}(2010)\citenamefont
  {Cherkasova}, \citenamefont {Tarasevich}, \citenamefont {Lebovka},\ and\
  \citenamefont {Vygornitskii}}]{Cherkasova2010}%
  \BibitemOpen
  \bibfield  {author} {\bibinfo {author} {\bibfnamefont {V.~A.}\ \bibnamefont
  {Cherkasova}}, \bibinfo {author} {\bibfnamefont {Y.~Y.}\ \bibnamefont
  {Tarasevich}}, \bibinfo {author} {\bibfnamefont {N.~I.}\ \bibnamefont
  {Lebovka}},\ and\ \bibinfo {author} {\bibfnamefont {N.~V.}\ \bibnamefont
  {Vygornitskii}},\ }\bibfield  {title} {\bibinfo {title} {Percolation of
  aligned dimers on a square lattice},\ }\href
  {https://doi.org/10.1140/epjb/e2010-00089-2} {\bibfield  {journal} {\bibinfo
  {journal} {The European Physical Journal B}\ }\textbf {\bibinfo {volume}
  {74}},\ \bibinfo {pages} {205} (\bibinfo {year} {2010})}\BibitemShut
  {NoStop}%
\bibitem [{\citenamefont {Du}\ \emph {et~al.}(2005)\citenamefont {Du},
  \citenamefont {Fischer},\ and\ \citenamefont {Winey}}]{Du2005}%
  \BibitemOpen
  \bibfield  {author} {\bibinfo {author} {\bibfnamefont {F.}~\bibnamefont
  {Du}}, \bibinfo {author} {\bibfnamefont {J.~E.}\ \bibnamefont {Fischer}},\
  and\ \bibinfo {author} {\bibfnamefont {K.~I.}\ \bibnamefont {Winey}},\
  }\bibfield  {title} {\bibinfo {title} {Effect of nanotube alignment on
  percolation conductivity in carbon nanotube/polymer composites},\ }\href
  {https://doi.org/10.1103/PhysRevB.72.121404} {\bibfield  {journal} {\bibinfo
  {journal} {Phys. Rev. B}\ }\textbf {\bibinfo {volume} {72}},\ \bibinfo
  {pages} {121404(R)} (\bibinfo {year} {2005})}\BibitemShut {NoStop}%
\bibitem [{\citenamefont {Park}\ \emph {et~al.}(2006)\citenamefont {Park},
  \citenamefont {Wilkinson}, \citenamefont {Banda}, \citenamefont {Ounaies},
  \citenamefont {Wise}, \citenamefont {Sauti}, \citenamefont {Lillehei},\ and\
  \citenamefont {Harrison}}]{Park2006}%
  \BibitemOpen
  \bibfield  {author} {\bibinfo {author} {\bibfnamefont {C.}~\bibnamefont
  {Park}}, \bibinfo {author} {\bibfnamefont {J.}~\bibnamefont {Wilkinson}},
  \bibinfo {author} {\bibfnamefont {S.}~\bibnamefont {Banda}}, \bibinfo
  {author} {\bibfnamefont {Z.}~\bibnamefont {Ounaies}}, \bibinfo {author}
  {\bibfnamefont {K.~E.}\ \bibnamefont {Wise}}, \bibinfo {author}
  {\bibfnamefont {G.}~\bibnamefont {Sauti}}, \bibinfo {author} {\bibfnamefont
  {P.~T.}\ \bibnamefont {Lillehei}},\ and\ \bibinfo {author} {\bibfnamefont
  {J.~S.}\ \bibnamefont {Harrison}},\ }\bibfield  {title} {\bibinfo {title}
  {Aligned single-wall carbon nanotube polymer composites using an electric
  field},\ }\href {https://doi.org/https://doi.org/10.1002/polb.20823}
  {\bibfield  {journal} {\bibinfo  {journal} {Journal of Polymer Science Part
  B: Polymer Physics}\ }\textbf {\bibinfo {volume} {44}},\ \bibinfo {pages}
  {1751} (\bibinfo {year} {2006})}\BibitemShut {NoStop}%
\bibitem [{\citenamefont {Mills}\ \emph {et~al.}(1995)\citenamefont {Mills},
  \citenamefont {Jónsson},\ and\ \citenamefont {Schenter}}]{MILLS1995305}%
  \BibitemOpen
  \bibfield  {author} {\bibinfo {author} {\bibfnamefont {G.}~\bibnamefont
  {Mills}}, \bibinfo {author} {\bibfnamefont {H.}~\bibnamefont {Jónsson}},\
  and\ \bibinfo {author} {\bibfnamefont {G.~K.}\ \bibnamefont {Schenter}},\
  }\bibfield  {title} {\bibinfo {title} {Reversible work transition state
  theory: application to dissociative adsorption of hydrogen},\ }\href
  {https://doi.org/https://doi.org/10.1016/0039-6028(94)00731-4} {\bibfield
  {journal} {\bibinfo  {journal} {Surface Science}\ }\textbf {\bibinfo {volume}
  {324}},\ \bibinfo {pages} {305} (\bibinfo {year} {1995})}\BibitemShut
  {NoStop}%
\bibitem [{\citenamefont {Jónsson}\ \emph {et~al.}()\citenamefont {Jónsson},
  \citenamefont {Mills},\ and\ \citenamefont {Jacobsen}}]{NEB_book}%
  \BibitemOpen
  \bibfield  {author} {\bibinfo {author} {\bibfnamefont {H.}~\bibnamefont
  {Jónsson}}, \bibinfo {author} {\bibfnamefont {G.}~\bibnamefont {Mills}},\
  and\ \bibinfo {author} {\bibfnamefont {K.~W.}\ \bibnamefont {Jacobsen}},\
  }\bibinfo {title} {Nudged elastic band method for finding minimum energy
  paths of transitions},\ in\ \href
  {https://doi.org/10.1142/9789812839664_0016} {\emph {\bibinfo {booktitle}
  {Classical and Quantum Dynamics in Condensed Phase Simulations}}},\ pp.\
  \bibinfo {pages} {385--404}\BibitemShut {NoStop}%
\bibitem [{\citenamefont {Henkelman}\ \emph {et~al.}(2000)\citenamefont
  {Henkelman}, \citenamefont {Uberuaga},\ and\ \citenamefont
  {Jónsson}}]{Henkelman2000}%
  \BibitemOpen
  \bibfield  {author} {\bibinfo {author} {\bibfnamefont {G.}~\bibnamefont
  {Henkelman}}, \bibinfo {author} {\bibfnamefont {B.~P.}\ \bibnamefont
  {Uberuaga}},\ and\ \bibinfo {author} {\bibfnamefont {H.}~\bibnamefont
  {Jónsson}},\ }\bibfield  {title} {\bibinfo {title} {{A climbing image nudged
  elastic band method for finding saddle points and minimum energy paths}},\
  }\href {https://doi.org/10.1063/1.1329672} {\bibfield  {journal} {\bibinfo
  {journal} {The Journal of Chemical Physics}\ }\textbf {\bibinfo {volume}
  {113}},\ \bibinfo {pages} {9901} (\bibinfo {year} {2000})}\BibitemShut
  {NoStop}%
\bibitem [{\citenamefont {Henkelman}\ and\ \citenamefont
  {Jónsson}(2000)}]{Henkelman2000b}%
  \BibitemOpen
  \bibfield  {author} {\bibinfo {author} {\bibfnamefont {G.}~\bibnamefont
  {Henkelman}}\ and\ \bibinfo {author} {\bibfnamefont {H.}~\bibnamefont
  {Jónsson}},\ }\bibfield  {title} {\bibinfo {title} {{Improved tangent
  estimate in the nudged elastic band method for finding minimum energy paths
  and saddle points}},\ }\href {https://doi.org/10.1063/1.1323224} {\bibfield
  {journal} {\bibinfo  {journal} {The Journal of Chemical Physics}\ }\textbf
  {\bibinfo {volume} {113}},\ \bibinfo {pages} {9978} (\bibinfo {year}
  {2000})}\BibitemShut {NoStop}%
\end{thebibliography}%

\end{document}